\documentclass[manuscript, screen]{acmart}

\newcommand{\hl}[2]{\relax}
\usepackage{tabularx} 
\usepackage{booktabs} 
\usepackage[ruled]{algorithm2e} 
\usepackage[colorinlistoftodos]{todonotes}
\usepackage{verbatim}
\usepackage{graphicx}

\usepackage{balance}       
\usepackage[T1]{fontenc}   
\usepackage{txfonts}
\usepackage{mathptmx}
\usepackage{hyperref}
\usepackage{color}
\usepackage{booktabs}
\usepackage{textcomp}

\usepackage{microtype}        
\usepackage{ccicons}          

\usepackage{cuted}
\usepackage{capt-of}

\usepackage{listings}

\usepackage{multirow}
\usepackage[normalem]{ulem}

\usepackage{blindtext}
\usepackage{subcaption}

\usepackage{enumitem}
\usepackage{graphicx}
\usepackage{multicol}

\usepackage{appendix}

\SetAlFnt{\small}
\SetAlCapFnt{\small}
\SetAlCapNameFnt{\small}
\SetAlCapHSkip{0pt}
\IncMargin{-\parindent}

\newcommand{\ej}[1]{\hl{purple}{EJ: #1}}
\newcommand{\rj}[1]{\hl{blue}{RJ: #1}}

\newcommand{\shortquote}[1]{``\emph{#1}''}
\newcommand{\longquote}[1]{\vspace{-1pt}\begin{quote}``\emph{#1}''\end{quote}}
\newcommand{\theme}[1]{\vspace{-1pt}\subsubsection{#1} }

\def\rqSteps{\textbf{RQ1 - Steps}\xspace}
\def\rqProcess{\textbf{RQ2 - Process}\xspace}
\def\rqTools{\textbf{RQ3 - Tools}\xspace}

\usepackage{setspace}

\newcolumntype{H}{>{\setbox0=\hbox\bgroup}c<{\egroup}@{}}

\def\yes{$\checkmark$}
\def\no{---}

\def\<#1>{\codeid{#1}}
\newcommand{\codeid}[1]{\ifmmode{\mbox{\small\ttfamily{#1}}}\else{\small\ttfamily #1}\fi}
\newcommand{\codeidsmall}[1]{\ifmmode{\mbox{\smaller\ttfamily{#1}}}\else{\smaller\ttfamily #1}\fi}

\def\plaintitle{Hypothesis Formalization: Empirical Findings, Software Limitations, and Design Implications}

\def\emptyauthor{}
\def\plainkeywords{statistical analysis; linear modeling; end-user programming; end-user elicitation; domain-specific language}

\makeatletter
\def\url@leostyle{%
  \@ifundefined{selectfont}{
    \def\UrlFont{\sf}
  }{
    \def\UrlFont{\small\bf\ttfamily}
  }}
\makeatother
\def\pprw{8.5in}
\def\pprh{11in}

\definecolor{linkColor}{RGB}{6,125,233}
\hypersetup{%
  pdftitle={\plaintitle},
  pdfauthor={\emptyauthor},
  pdfkeywords={\plainkeywords},
  pdfdisplaydoctitle=true, 
  bookmarksnumbered,
  pdfstartview={FitH},
  colorlinks,
  citecolor=black,
  filecolor=black,
  linkcolor=black,
  urlcolor=linkColor,
  breaklinks=true,
  hypertexnames=false
}

\hypersetup{draft}
\begin{document}

\title{\plaintitle}

\begin{abstract}
Data analysis requires translating higher level questions and hypotheses
into computable statistical models. We present a mixed-methods study aimed
at identifying the steps, considerations, and challenges involved in
operationalizing hypotheses into statistical models, a process we refer to
as \textit{hypothesis formalization}. In a content analysis of  research papers, we
find that researchers highlight decomposing a hypothesis into
sub-hypotheses, selecting proxy variables, and formulating statistical
models based on data collection design as key steps. In a lab study, we find
that analysts fixated on implementation and shaped their analysis to fit
familiar approaches, even if sub-optimal. In an analysis of software tools,
we find that tools provide inconsistent, low-level abstractions that may
limit the statistical models analysts use to formalize hypotheses. Based on
these observations, we characterize hypothesis formalization as a
dual-search process balancing conceptual and statistical considerations
constrained by data and computation, and discuss implications for future tools.
\end{abstract}

\author{Eunice Jun}
\email{emjun@cs.washington.edu}
\affiliation{%
  \institution{University of Washington}
  \city{Seattle}
  \state{Washington}
  \country{USA}
}
\author{Melissa Birchfield}
\email{mbirch2@cs.washington.edu}
\affiliation{%
  \institution{University of Washington}
  \city{Seattle}
  \state{Washington}
  \country{USA}
}
\author{Nicole de Moura}
\email{nicoledemoura4@gmail.com}
\affiliation{%
  \institution{Eastlake High School}
  \city{Sammamish}
  \state{Washington}
  \country{USA}
}
\author{Jeffrey Heer}
\email{jheer@cs.washington.edu}
\affiliation{%
  \institution{University of Washington}
  \city{Seattle}
  \state{Washington}
  \country{USA}
}
\author{Ren{\'e} Just}
\email{rjust@cs.washington.edu}
\affiliation{%
  \institution{University of Washington}
  \city{Seattle}
  \state{Washington}
  \country{USA}
}
\maketitle

\ccsdesc[500]{Human-centered computing~User interface toolkits}

\keywords{\plainkeywords}


\newcommand{\formula}{Formula}
\newcommand{\assignRoles}{Specify variable roles}

\def\arraystretch{0.35}
\newcommand{\tableSoftwareAnalysis}{
    \begin{table*}
        \footnotesize
        \caption{\textbf{Overview of the software tools included in our analysis.}\label{tableAnalysisOfTools}}
        \begin{small}
        \vspace*{-10pt}
        \begin{minipage}{\linewidth}
        Half of the tools
        are specialized for specific modeling use cases. Most tools
        use mathematical notation (T18--T20 (\yes*) even use
        mathematical notation in their GUIs).
        Most tools also provide a wide range
        of computational control although sometimes they require additional
        packages [T5, T13]. Tool specialization, organization,
        notation, and computational control focus analysts on model
        implementation details, sometimes at the expense of focusing on their conceptual
        hypotheses.
        \end{minipage}
        \end{small}

        \setlength{\tabcolsep}{4pt}
        \begin{tabular}{l>{\raggedright}p{0.3\linewidth}p{0.1\linewidth}p{0.09\linewidth}p{0.1\linewidth}p{0.1\linewidth}}
        \toprule
        ID & Tool name                                          & Specialized Scope    & Mathematical Notation        & Computational Control            & References                               \\                     
        \multicolumn{2}{l}{\textbf{R Packages}} \\
        \midrule\\
        T1 & MASS                                               & \no                  & \yes                         & \yes                             & ~\cite{mass}                                 \\                     
        T2 & brms                                               & \yes                 & \yes                         & \yes                             & ~\cite{burkner2017brms,brmsRef}                                 \\                     
        T3 & edgeR                                              & \yes                  & \yes                        & \yes                            & ~\cite{edgeROverview,edgeRUsersGuide}                                 \\                     
        T4 & glmmTMB                                            & \yes                 & \yes                         & \yes                             & ~\cite{glmmtmbPaper,glmmtmbRef}                                 \\                     
        T5 & glmnet                                             & \yes                 & \no                          & \yes (additional)                & ~\cite{glmnetRef,glmnetVignette}                                 \\                     
        T6 & lme4                                               & \yes                 & \yes                         & \yes                             & ~\cite{bates2014fittingLme4,bates2014lme4Ref}                                 \\                     
        T7 & MCMCglmm                                           & \yes                 & \yes                         & \yes                             & ~\cite{MCMCglmmPaper,MCMCglmmRef}                                 \\                     
        T8 & nlme                                               & \yes                 & \yes                         & \yes                             & ~\cite{nlmeRef}                                 \\                     
        T9 & RandomForest                                       & \yes                 & \yes                         & \yes (minimal)                   & ~\cite{randomForestR}                                 \\                     
        T10 & stats (core library)                              & \no                  & \yes                         & \yes                             & ~\cite{statsCoreRRef}                                 \\                     
        \multicolumn{2}{l}{\textbf{Python Packages}} \\                 
        \midrule\\                  
        T11 & Keras                                             & \yes                 & \no                          & \yes (minimal)                   & ~\cite{keras}                                 \\                     
        T12 & Scikit-learn                                      & \yes                 & \no                          & \yes                             & ~\cite{scikitRef,scikitPaper,scikitAPIPaper}                                 \\                                                                            
        T13 & Scipy (scipy.stats)                               & \no                  & \no                          & \yes (additional)                & ~\cite{scipy,scipyStats,scipyOptimize}                                 \\                     
        T14 & Statsmodels                                       & \no                  & \yes                         & \no                              & ~\cite{statsmodelsPaper,statsmodelsRef}                                 \\                                                                                  
        \multicolumn{2}{l}{\textbf{Suites, with DSLs for programming}} \\                   
        \midrule\\                  
        T15 & Matlab (Statistics and ML Toolbox)                & \no                  & \no                          & \yes                             & ~\cite{matlab,matlabStats}                                 \\                     
        T16 & SPSS                                              & \no                  & \yes                         & \yes                             & ~\cite{spss}                                 \\                                    
        T17 & Stata                                             & \no                  & \yes                         & \no                              & ~\cite{stata,stataRef,stataLang}                                 \\                     
        \multicolumn{2}{l}{\textbf{Suites, without programming}} \\                 
        \midrule\\                  
        T18 & GraphPrism                                        & \no                  & \yes *                       & \yes                             & ~\cite{graphPadUserGuide}                                 \\                                    
        T19 & JASP                                              & \no                  & \yes *                       & \no                              & ~\cite{jasp}                                 \\                                    
        T20 & JMP                                               & \no                  & \yes *                       & \no                              & ~\cite{jmp,jones2011jmp}                                 \\                                                                                  
        \bottomrule 
        \end{tabular}
        \end{table*}
}

\newcommand{\tableLitSurvey}{
\begin{table} [hb!]
    \small
    \centering
    \caption{\textbf{Overview of research article sample.}}
    \begin{tabular}{p{0.4\columnwidth}p{0.3\columnwidth}p{0.2\columnwidth}} 
    \textbf{Venue} & \textbf{Articles} & \textbf{Sample Size} \\
    \hline
    Psychological Science & \cite{perfecto2019volume,oh2019revealing,cao2019people,jefferies2019sudden,nave2019bigger,golle2019school,jouravlev2019tracking} & \\
    Nature & \cite{liu2016deletions,herault2017myeloid,iaccarino2016gamma,huang2016structural,kremkow2016principles,dilley2016break} & \\
    CHI &  & \\
    & & Total: \\
    \end{tabular}
    \label{table:litSurvey}
\end{table}
}

\newcommand{\figureFlow}{
    \begin{figure}[t]
        \centering
        \includegraphics[width=0.5\textwidth]{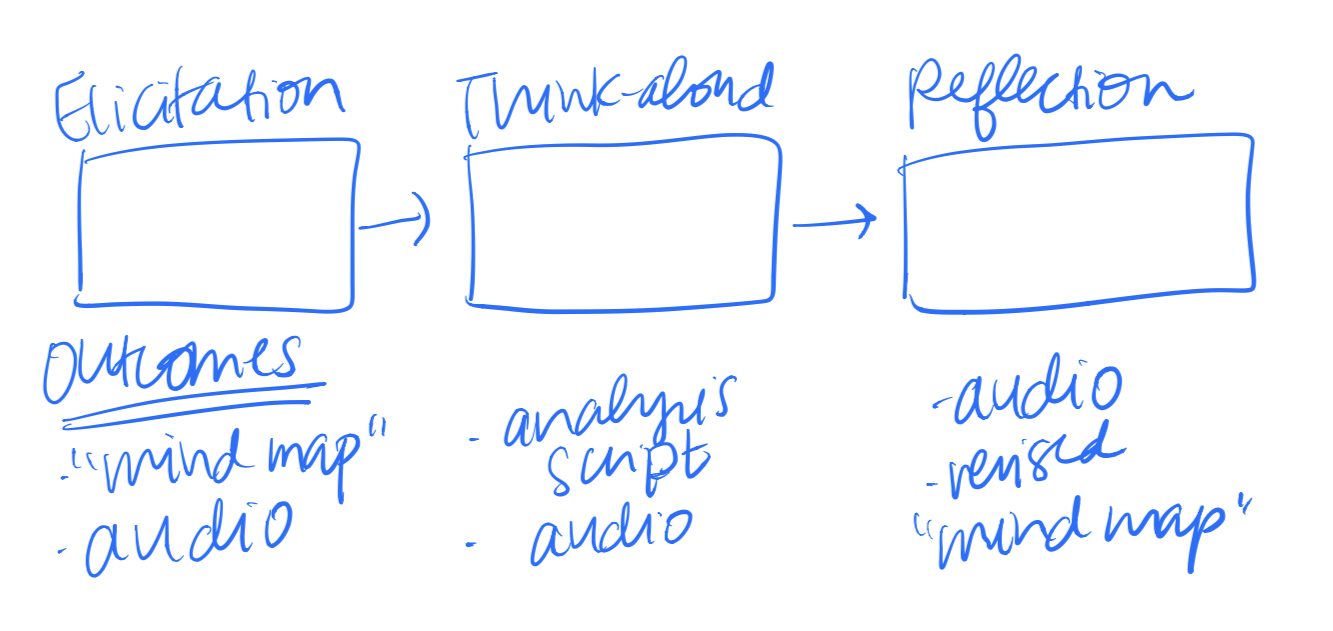}
        \caption{Proposed flow for in-lab study. There will be three different parts. From left to right: natural programming elicitation, think-aloud analysis, and reflective interview. The outcomes for each stage are listed below each stage.}
        \label{figure:flow}
      \end{figure}
}

\newcommand{\figureOverivew}{
    \begin{figure}[t]
        \centering
        \includegraphics[width=.9\textwidth]{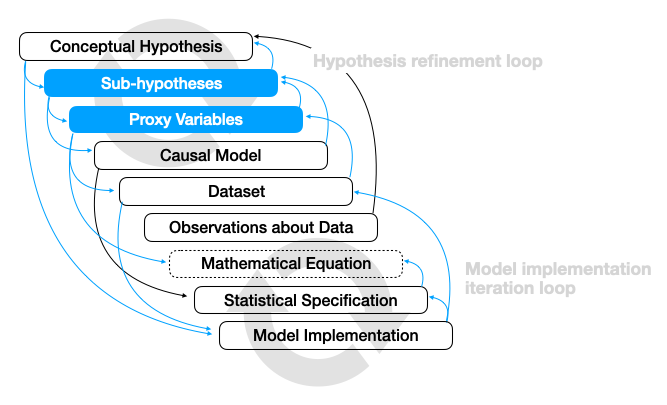}
        \caption{\textbf{Definition and overview of the hypothesis formalization
        steps and process.}\label{figure:overview}}
        \begin{small}
        \begin{minipage}{\linewidth}
        Hypothesis formalization is a dual-search process of
        translating a \textbf{conceptual hypothesis} into a statistical
        \textbf{model implementation}.
        Blue indicates steps and transitions that we identified.
        Black indicates steps and transitions discussed in prior work.
        ``Mathematical Equation'' (dashed box) was rarely an explicit step in
        our lab study but evident in our content analysis.
        Our findings (blue arrows) subsume several of the transitions identified
        in prior work. When they do not, prior work's transitions are included in black.
        Hypothesis formalization is a non-linear process.
        Analysts iterate over conceptual steps to refine their hypothesis in a
        \textit{hypothesis refinement loop}.
        Analysts also iterate over computational and implementation steps in a
        \textit{model implementation loop}.
        Data collection and data properties may also prompt conceptual revisions
        and influence statistical model implementation.
        As analysts move toward model implementation, they increasingly rely on
        software tools, gain specificity, and create intermediate artifacts
        along the way (e.g., causal models, observations about data, etc.).
        \end{minipage}
        \end{small}
      \end{figure}
}

\newcommand{\figurePriorWork}{
    \begin{figure}[t]
        \centering
        \includegraphics[width=0.7\textwidth]{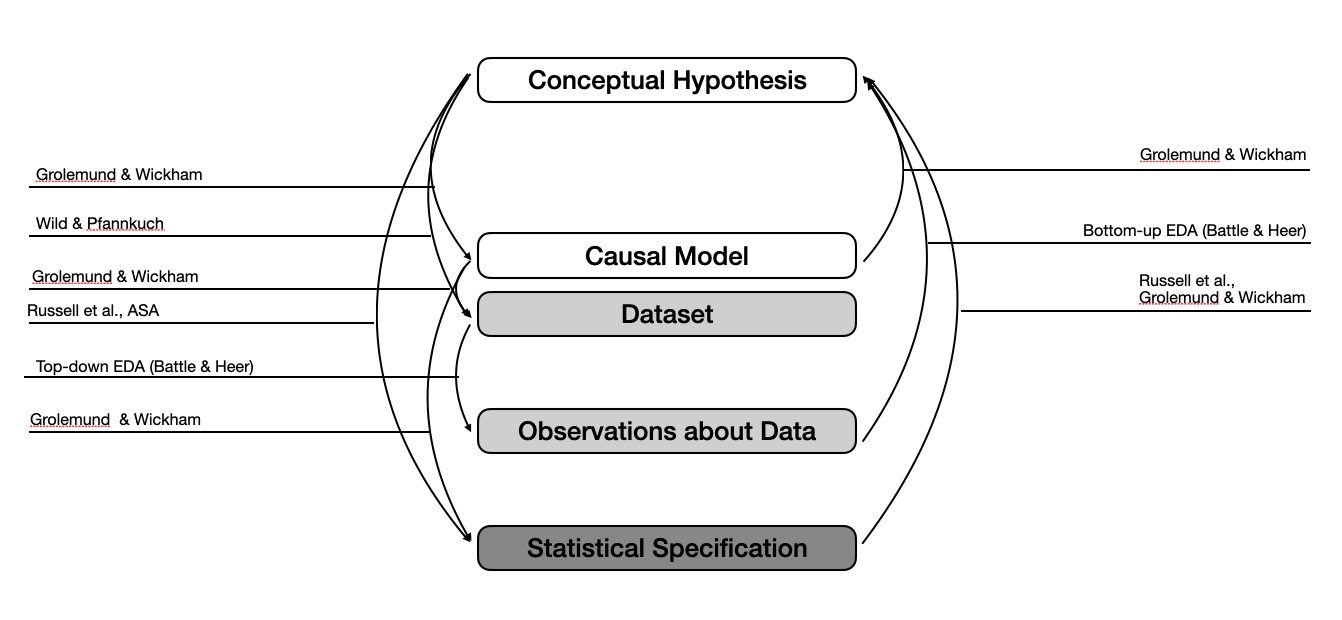}
        \caption{\textbf{Expected steps based on prior work.} Prior work on scientific discovery, theories of sensemaking, statistical thinking, and empirical studies of analysis suggest the above series of steps involved in hypothesis formalization. We find more granular steps and more back and forth between steps involved in hypothesis formalization, as shown in Figure 1.}
        \label{figure:priorWork}
      \end{figure}
}

\newcommand{\figureMethods}{
    \begin{figure}[t]
        \centering
        \includegraphics[width=0.7\textwidth]{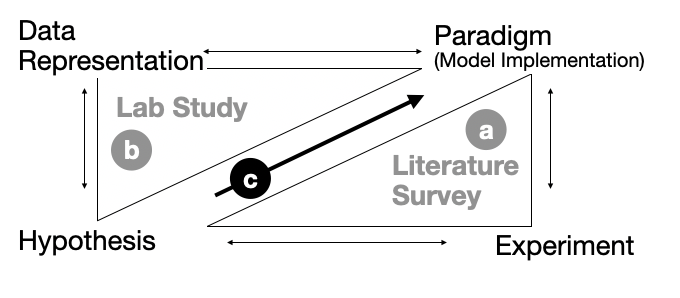}
        \caption{\textbf{Methods in the context of scientific discovery.} Hypothesis formalization is a part of the scientific discovery process, which Schunn and Klahr describe as involving four spaces~\cite{schunn1995FourSpace}, as originally modeled as above. (a) In the content analysis of research publications (\autoref{sec:litSurvey}), we touch on the hypothesis, paradigm (which includes model implementation), and experiment spaces. (b) The lab study (\autoref{sec:inLabStudy}) gives us insight into the hypothesis, paradigm, and data representation spaces. (c) Together with the heuristic analysis of tools (\autoref{sec:toolsAnalysis}), these methods provide us with complementary perspectives on the transition from hypothesis to paradigm spaces, which is the focus of hypothesis formalization and this paper. Although Schunn and Klahr's original model only includes information flowing from hypotheses to paradigms, we find that hypothesis formalization involves alternation between the two spaces. \ej{This finding highlights the tension between linear and iterative processes involved in hypothesis formalization.}}
        \label{figure:methods}
      \end{figure}
}

\newcommand{\figurePriorWorkCombined}{
    \begin{figure}
        \centering
        \includegraphics[width=0.8\textwidth]{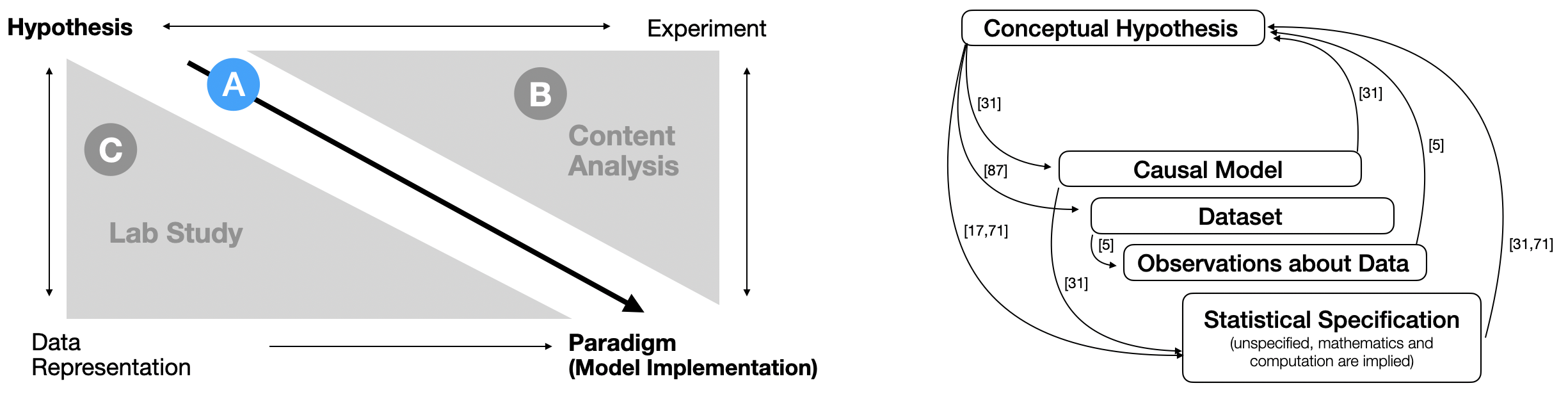}
        \caption{\textbf{Relationship between hypothesis formalization and prior work.}\label{figure:priorWork}}
        \begin{small}
        \begin{minipage}{\linewidth}
         \textit{Left:}
        Schunn and
        Klahr's four-space model of scientific
        discovery (stylized adaptation from  Figure 1 in~\cite{schunn1995FourSpace}),
        which includes unidirectional information flow
        from the hypothesis space to the paradigm space (which includes
        model implementation). Hypothesis formalization (A) is focused on a
        tighter integration and the information flow between hypothesis and paradigm spaces.
        Specifically, the information flow is bidirectional in hypothesis formalization.
        Our content analysis (B) and lab study (C) triangulate the four-space
        model to understand hypothesis formalization
        from complementary perspectives.
        \textit{Right:} Hypothesis formalization steps also identified in
        prior work on theories of
        sensemaking, statistical thinking, and data analysis workflows (citations included to the right of the arrows).
        Hypothesis formalization is finer grained and involves more iterations.
        While prior work broadly refers to mathematical equations,
        partial model specifications, and computationally tuned model
        implementations as statistical specifications, hypothesis formalization
        differentiates them. This paper provides empirical evidence for
        theorized loops between conceptual hypothesis and statistical
        specification (see Figure~\ref{figure:overview}).
        \end{minipage}
        \end{small}
        \vspace*{-15pt}
      \end{figure}
}

\newcommand{\figureImplications}{
    \begin{figure}[t]
        \centering
        \includegraphics[width=\textwidth]{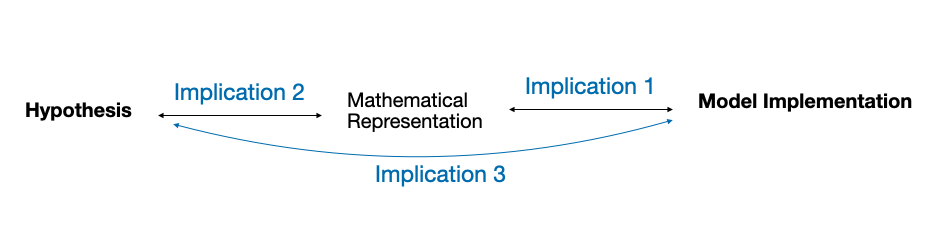}
        \caption{\textbf{Implications for Future Tools.} Our characterization of hypothesis formalization suggests three implications for tightening the paths from hypotheses to model implementations, the two end-points in hypothesis formalization. Implication 1 aims to strengthen the connection between model implementations and their underlying mathematical representations. Implication 2 builds on Implication 1 to translate hypotheses into mathematical representations and model implementations. Implication 3 aims to support the back and forth between hypotheses and model implementations in a way that may not involve mathematical representations.}
        \label{figure:implications}
      \end{figure}
}

\newcommand{\tableparticipants}{
    \begin{table}
        \small
        \singlespacing
        \centering
        \caption{\textbf{Overview of participants.} In a pre-study survey with questions modeled after~\cite{},
        participants self-reported having a median experience of 7 with data
        analysis and of 6.5 with programming on a scale of 1, very
        inexperienced, to 10, very experienced. Participants also came from different disciplinary backgrounds and jobs. Of the 24 recruited participants, 18 completed all three phases of the study. }
        \begin{tabular}{lp{0.2\textwidth}p{0.2\textwidth}p{0.1\textwidth}p{0.1\textwidth}l}
        PID & Domain                            & Job role                  & Analysis                         & Programming                  & Completed    \\ 
            &                                   &                           & Experience                       & Experience                   & Study    \\                    
        \hline
        1 & Computer Science                    & Graduate student          & 6                               & 8                             &  \yes  \\ 
        2 & Data Science                        & Freelancer                & 7                               & 8                             &  \yes  \\ 
        3 & Data science                        & Freelancer                & 8                               & 8                             &  \yes   \\ 
        4 & Psychology                          & Graduate student          & 6                               & 4                             &  \yes \\ 
        5 & Other - Narrative design            & Unemployed                & 2                               & 2                             &  \yes  \\ 
        6 & Psychology                          & Research Scientist        & 8                               & 4                             &  \yes   \\ 
        7 & Unspecified                         & Recent Master's graduate  & 7                               & 7                             &  \yes   \\ 
        8 & Data science                        & Analyst, Consultant       & 7                               & 6                             &  \no   \\ 
        9 & Biology                             & Graduate student          & 7                               & 6                             &  \no  \\ 
        10& Business, Marketing                 & Analyst                   & 10                              & 7                             &  \yes   \\ 
        11& Biology                             & Scientist                 & 7                               & 7                             &  \yes   \\ 
        12& Data science                        & Project Manager           & 4                               & 3                             &  \no   \\ 
        13& Data science                        & Recent undergraduate      & 6                               & 3                             &  \yes  \\ 
        14& Data science                        & Recent undergraduate      & 3                               & 3                             &  \yes   \\ 
        15& Data science                        & Analyst                   & 7                               & 8                             &  \yes   \\ 
        16& Business                            & Recent MBA graduate       & 9                               & 5                             &  \yes  \\ 
        17& Statistics                          & Freelancer, Consultant    & 3                               & 3                             &  \no   \\ 
        18& Computer Science                    & Student (unspecified)     & 6                               & 8                             &  \yes   \\ 
        19& Education                           & Educator                  & 6                               & 7                             &  \no   \\ 
        20& Computer Science                    & Recent undergraduate      & 8                               & 9                             &  \yes   \\ 
        21& Healthcare                          & Software Engineer         & 4                               & 8                             &  \yes   \\ 
        22& Neuroscience                        & Analyst                   & 8                               & 5                             &  \no   \\ 
        23& Biology                             & Undergraduate student     & 5                               & 3                             &  \yes   \\ 
        24& Physics                             & Graduate student          & 9                               & 9                             &  \yes     
        \end{tabular}
        \label{table:participants}
        \end{table}
}

\newcommand{\tableHypotheses}{
    \begin{table}
        \small
        \centering      
        \caption{\textbf{Overview of hypotheses generated in the lab study.}} 
        \begin{tabular}{llll}
            Form & Examples & Total Instances & Participants \\
            \hline \\
        \end{tabular}
        \label{table:hypotheses}
    \end{table}
}

\newcommand{\tableConceptualModels}{
    \begin{table}
        \small
        \centering      
        \caption{\textbf{Overview of conceptual models participants developed in the lab study.}\ej{Add the occurrence numbers once decide on metric. Consider removing the examples to reduce space. Or move entirely to appendix?}} 
        \begin{tabular}{llll}
            Form & Examples & Total Instances & Participants \\
            \hline \\
        \end{tabular}
        \label{table:conceptualModels}
    \end{table}
}

\newcommand{\figureLitSurveyMatrix}{
    \begin{figure}
        \centering
        \includegraphics[width=1.\textwidth]{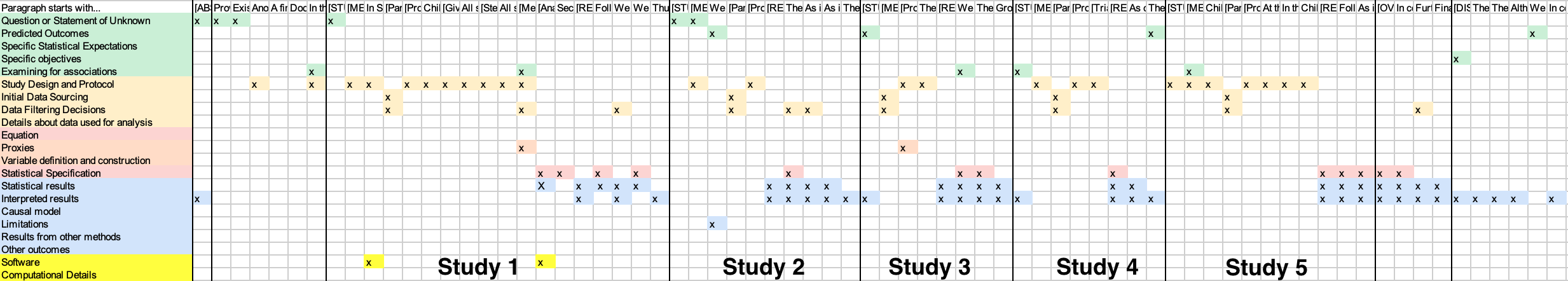}
        \caption{\textbf{Reorderable Matrix for a Psychological Science paper in our literature survey.} The paper reports on five studies. The black vertical lines designate distinctions between the abstract, introduction, five studies, and conclusion to make the paper structure more obvious. Notice the clusters that are noticeable for each of the five studies. Also notice that the research goals start broad and become more specific for each study. \ej{Replace the figure with a polished one annotated with x- and y-axes. And maybe some callouts for specific studies and hypotheses since the figure is so wide.}}
        \label{figure:litSurveyMatrix}
    \end{figure}
}

\newcommand{\figureLabStudyStatSpec}{
    \begin{figure}
        \centering
        \begin{minipage}{0.45\textwidth}
            \centering
            \includegraphics[width=0.9\textwidth]{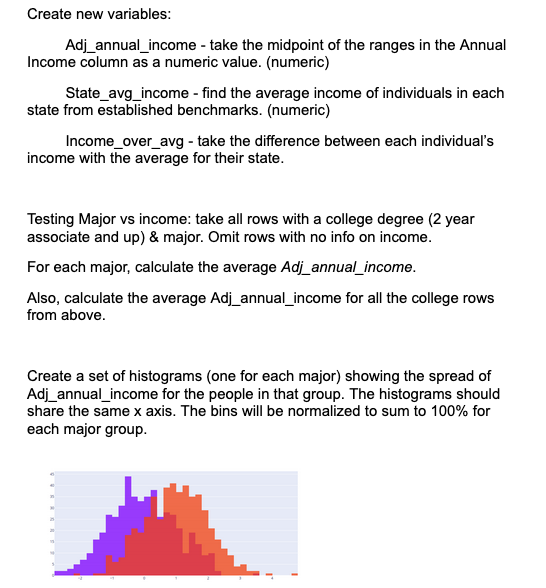} 
        \end{minipage}\hfill
        \begin{minipage}{0.45\textwidth}
            \centering
            \includegraphics[width=0.9\textwidth]{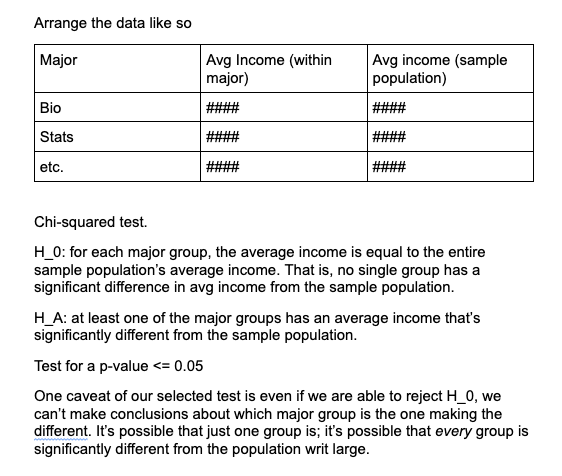} 
        \end{minipage}
        \caption{\textbf{Sample statistical specification (A8).}\label{figure:labStudyStatSpec}}
            \begin{small}
            \begin{minipage}{\linewidth}
            The lab study tasked analysts
            to specify their statistical models without
            considering implementation.
            Analyst A8 wrote a list (split across two pages) of detailed steps
            involved in creating new variables, cleaning and wrangling data,
            visualizing data, and testing their hypothesis.
            We expected analysts would articulate
            their statistical models mathematically. Instead, analysts specified
            their statistical models using todo lists and summaries of steps,
            which sometimes included mentions of software tools, showing that
            implementation was an important consideration and
            that tool familiarity may limit which statistical models
            analysts consider and implement.
            \end{minipage}
            \end{small}
    \end{figure}
}

\figureOverivew 

\section{Introduction}
Using statistics to answer real-world questions requires four steps: (i) translating
high-level, domain-specific questions and hypotheses into specific statistical
questions~\cite{carver2016guidelines}; (ii) identifying statistical
models to answer the statistical questions; (iii) implementing
and executing these statistical models, typically with the help of software tools;
and (iv) interpreting the results, considering
the domain-specific questions and applying analytical reasoning.



For example, suppose a census researcher asked, ``In the United States (U.S.), how does
an individual's sex relate to their annual income?'' Drawing upon their prior
experiences and exploratory data visualizations, the researcher knows that
income in the U.S. is skewed, and they want to know how the distributions of
income among males and females differ (step i). However, before implementing,
they (implicitly) define their causal model: The researcher knows that
other factors, such as education and race, may be associated with employment
opportunities, which may then influence income. As such, they refine their
conceptual hypothesis\rj{What was the conceptual hypothesis to begin with?} to consider the possible effects of race, education, sex, and their
interactions on income. They plan to fit a generalized linear model with
race, education, sex, and their two-way interactions as predictors of income
(step ii). They start implementing a script to load and model
data (step iii). The researcher receives a small table of results and is
surprised to receive a convergence warning\rj{It's not clear to me why the
described model would not converge (e.g., perfect separation seems unrealistic)}. After further investigation,
they simplify their model and remove the interaction effects to see how
that may affect convergence (revise step iii). This time, their model converges,
and they interpret the results (step iv), but they really want to study how
sex and race interact, so they return to implementation (step iii) and proceed
as before, iteratively removing and adding effects and changing computational
parameters, and as a by-product shifting which high-level conceptual hypothesis
is reflected in the model.


Performing statistical data analysis goes well beyond invoking the correct statistical functions in a software library.
Analysts, such as the census researcher, must go back and forth
between conceptual hypothesis and model implementation realities, grappling with
domain knowledge, limitations of data, and statistical methods.

We refer to the process of translating a conceptual hypothesis into a computable statistical
model as \textit{hypothesis formalization}. This process is  messy and under-scrutinized, and consequently
we investigate the steps, considerations, challenges, and tools involved.
Based on our findings, we
define hypothesis formalization as a dual-search process~\cite{klahr1988dual}
that involves developing and integrating cognitive representations from two
different perspectives---conceptual hypotheses and concrete model
implementations. Analysts move back and forth between these two perspectives
during formalization while balancing conceptual,
data-driven, statistical, and implementation constraints.
Figure~\ref{figure:overview} summarizes our definition and findings. 
Specifically, the paper addresses the following questions:
\begin{itemize}
    \item \rqSteps: What is the range of steps an analyst might consider when formalizing a hypothesis? How do these steps compare to ones that we might expect based on prior work?
    \item \rqProcess: How do analysts think about and perform the steps to translate their hypotheses into model implementations? What challenges do they face during this process?
    \item \rqTools: How might current software tools influence hypothesis formalization?
\end{itemize}

To develop a broad understanding of the steps (\rqSteps) and considerations
(\rqProcess) involved in hypothesis formalization, we first conducted a content
analysis of 50 randomly sampled research papers across five different domains. We
find that researchers decompose their research hypotheses into specific sub-hypotheses,
derive proxy variables from theory and available data, and
adapt statistical analyses to account for data collection procedures.

To validate and deepen our understanding of
hypothesis formalization (\rqSteps and \rqProcess), we designed and conducted a lab
study in which we observed 24 analysts develop and formalize hypotheses in-situ. We
find that analysts foreground implementation concerns, even when
brainstorming hypotheses, and try to fit their hypotheses and
analyses to prior experiences and familiar tools, suggesting a strong influence of
tools (\rqTools). 


To identify how tools may shape hypothesis formalization (\rqTools), we reviewed
20 statistical software tools. We find that although the tools support nuanced
model implementations, their low-level abstractions can focus analysts on
statistical and computational details at the expense of higher-level reasoning
about initial hypotheses. Tools also do not aid analysts in
identifying reasonable model implementations that would test their conceptual
hypotheses, which may explain why analysts in our lab study relied on familiar
approaches, even if sub-optimal.

Our content analysis, lab study, and tools analysis inform one another
and suggest \textbf{three design implications} for tools to more directly support
hypothesis formalization: (i) connect statistical model
implementations to their mathematical representations\rj{I don't understand this point.}, (ii) provide higher-level
abstractions for expressing conceptual hypotheses and partial model
specifications, and (iii) develop bidirectional computational assistance for
authoring causal models and relating them to statistical models.

By defining and characterizing hypothesis formalization, we aim to advance the
discourse around data analysis, more precisely understand how people
conduct data analysis, reveal opportunities for more transparent
and reproducible analyses, and inform future tools
that support end-users in their analysis processes.

\section{Background and Related Work}
Our work integrates and builds upon prior research on frameworks of scientific
discovery, theories of sensemaking, statistical practices, and empirical studies
of data analysts. 

\subsection{Dual-search Model of Scientific Discovery}
Klahr and Simon characterized scientific discovery as a dual-search process
involving the development and evaluation of hypotheses and
experiments~\cite{klahr1988dual}. They posited that scientific
discovery involved tasks specific to hypotheses (e.g., revising hypotheses) and
to experiments (e.g., analyzing data collected from experiments), which they
separated into two different ``spaces,'' and tasks moving between them, which is
where we place hypothesis formalization.

Extending Klahr and Simon's two-space model, Schunn and Klahr proposed a more
granular four-space model involving data representation, hypothesis, paradigm,
and experiment spaces~\cite{schunn1995FourSpace,schunn1996BeyondTwoSpace}. In the four-space model, conceptual hypothesizing still lies in the
hypothesis space, and hypothesis testing and statistical modeling lies in the
paradigm space. As such, hypothesis formalization is a process connecting
the hypothesis and paradigm spaces. In Schunn and Klahr's four-space model,
information flows unidirectionally from the hypothesis space to the paradigm space.
Here we extend this prior research with
evidence that hypothesis formalization involves both
concept-to-implementation and implementation-to-concept processes. (see
Figure~\ref{figure:overview}).
Figure~\ref{figure:priorWork} augments Schunn and Klahr's
original diagram (Figure 1 in~\cite{schunn1995FourSpace}) with
annotations depicting how our content analysis of research papers and lab study
triangulate a tighter dual-space search between hypothesis and
paradigm spaces with a focus on hypothesis formalization. Our mixed-methods
approach follows the precedent and recommendations of Klahr and
Simon's~\cite{klahr1999studies} study of scientific discovery activities.

\figurePriorWorkCombined

\subsection{Theories of Sensemaking}
Human beings engage in \textit{sensemaking} to acquire new knowledge. Several
theories of
sensemaking~\cite{pirolli2005sensemaking,russell1993cost,klein2007dataFrame}
describe how and when human beings seek and integrate new data (e.g.,
observations, experiences, etc.) to develop their mental models about the world.

Russell et al.~\cite{russell1993cost} emphasize the importance of building up
and evaluating external representations of mental models, and define sensemaking
as ``the process of searching for a representation and encoding data in that
representation to answer task-specific questions.'' External representations are
critical because they influence the quality of conclusions reached at the end of
the sensemaking process and affect how much time and effort is required in the process. Some representations
may lead to insights more quickly. Russell et al. describe the iterative
process of searching for and refining external representations in a ``learning
loop complex'' that involves transitioning back and forth between (i) searching
for and (ii) instantiating representations. 
 
Grolemund and Wickham argued for statistical data analysis as a sensemaking
activity~\cite{grolemund2014cognitive}. They emphasize the (1)
bidirectional nature of updating mental models of the world and hypotheses based
on data and collecting data based on hypotheses and (2) the process of
identifying and reconciling discrepancies between hypotheses and data. Their
depiction of the analysis process parallels Klahr and Simon's framework of
scientific discovery.


In this paper, we consider hypothesis formalization to be a learning loop~\cite{russell1993cost} where
the conceptual hypothesis is an external representation of a set of assumptions
analysts may have about the world (e.g., an implicit causal model), that ultimately
affects which models are specified and which results are
obtained. We found that that there are smaller learning loops as analysts search
for and revise intermediate representations, such as explicit causal models,
mathematical equations, or partially specified models. The
hypothesis and model refinement loops can themselves be smaller learning loops
embedded in the larger loop of hypothesis formalization. 


\subsection{Statistical Thinking} 
Statistical thinking and practice require differentiating between
\textit{domain} and \textit{statistical} questions. The American Statistical
Association (ASA), a professional body representing statisticians, recommends
that universities teach this fundamental principle in introductory courses (see
Goal 2 in~\cite{carver2016guidelines}). 

Similarly, researchers Wild and Pfannkuch emphasize the importance of
differentiating between and integrating statistical knowledge and context (or
domain) knowledge when thinking
statistically~\cite{pfannkuch1997statistical,pfannkuch2000statistical,wild1999statisticalThinking}.
They propose a four step model for operationalizing ideas (``inklings'') into
plans for collecting data, which are eventually statistically analyzed. In their
model, analysts must transform ``inklings'' into broad questions and then into
precise questions that are then finally turned into a plan for data collection
(see Figure 2 in~\cite{wild1999statisticalThinking}). Statistical and domain
knowledge inform all four stages. However, it is unknown what kinds of statistical and domain
knowledge are helpful, how they are used and weighed against each other, and
when certain kinds of knowledge are helpful to operationalize inklings. Our work provides more granular insight into Wild and Pfannkuch's
proposed model of operationalization and aims to answer when, how, and what
kinds of statistical and domain knowledge are used during statistical data
analysis. 

More recently, in \textit{Statistical
Rethinking}~\cite{mcelreath2020statistical}, McElreath proposes that
there are three key representational phases involved in data analysis:
conceptual hypotheses, causal models underlying hypotheses (which McElreath
calls ``process models''), and statistical models. McElreath, like the ASA and
Wild and Pfannkuch, separates domain and statistical ideas and discusses the use
of causal models as an intermediate representation to connect
the two. McElreath emphasizes that conceptual hypotheses may correspond to
multiple causal and statistical models, and that the same statistical
model may provide evidence for multiple, even contradictory, causal models and
hypotheses. McElreath's framework does not directly address how analysts navigate
these relationships or how computation plays a role, both of which we take up in
this paper. 

Overall, our work provides empirical evidence for prior frameworks but also (i)
provides more granular insight into \textit{how} and \textit{why} transitions between
representations occur and (ii) scrutinizes the role of
\textit{software and computation} through close observation of analyst workflows
in the lab as well as through a follow-up analysis of statistical software. Based on
these observations, we also speculate on how tools might better support hypothesis
formalization.
\subsection{Empirical Studies of Data Analysts}
Data analysis involves a number of tasks that involve data discovery, wrangling,
profiling, modeling, and reporting~\cite{kandel2012enterprise}. Extending the findings of
Kandel et al., both Alspaugh et al.~\cite{alspaugh2018futzing} and
Wongsuphasawat et al.~\cite{wongsuphasawat2019EDAgoals}
propose exploration as a distinct task.
Whereas Wongsuphasawat et al. argue that exploration should subsume
discovery and profiling, Alspaugh et al. describe exploration as an alternative
to modeling. The importance of exploration and its role in updating analysts'
understanding of the data and their goals and hypotheses is of note, regardless
of the precise order or set of tasks. Battle and Heer describe exploratory
visual analysis (EVA), a subset of exploratory data analysis (EDA) where
visualizations are the primary outputs and interfaces for exploring data, as
encompassing both data-focused (bottom-up) and goal- or hypothesis-focused
(top-down) investigations~\cite{battle2019EVA}. In our lab study, we found that
(i) analysts explored their data before modeling and (ii) exploratory
observations sometimes prompted conceptual shifts in hypotheses (bottom-up) but
at other times were guided by hypotheses and only impacted statistical
analyses (top-down). In this way, data exploration appears to be an important
intermediate step in hypothesis formalization, blurring the lines between
exploratory and confirmatory data analysis. 

Decisions throughout analysis tasks can give rise to a ``garden of forking
paths''~\cite{gelman2013garden}, which compounds for meta-analyses synthesizing
previous findings~\cite{kale2019decision}. Liu, Boukhelifa, and
Eagan~\cite{liu2019understanding} proposed a broad framework that characterizes
analysis alternatives using three different \textit{levels of abstraction}:
cognitive, artifact, and execution. \textit{Cognitive} alternatives involve more
conceptual shifts and changes (e.g., mental models, hypotheses).
\textit{Artifact} alternatives pertain to tooling (e.g., which software is used
for analysis?), model (e.g., what is the general mathematical approach?), and
data choices (e.g., which dataset is used?). \textit{Execution} alternatives are
closely related to artifact alternatives but are more fine-grained programmatic
decisions (e.g., hyperparameter tuning). We find that hypothesis formalization
involves all three levels of abstraction. We provide a more granular depiction
of how these levels cooperate with one another. 

Moreover, Liu, Althoff, and Heer~\cite{liu2019paths} identified numerous
decision points throughout the data lifecycle, which they call
\textit{end-to-end analysis}. They found that analysts often revisit key
decisions during data collection, wrangling, modeling, and evaluation. Liu,
Althoff, and Heer also found that researchers executed and selectively reported
analyses that were already found in prior work and familiar to the research
community. Hypothesis formalization is comprised of a subset of steps involved
in end-to-end analysis. Thus, we expect hypothesis formalization will be an
iterative process where domain norms will influence decision making. It is
nonetheless valuable to provide insight into how a single iteration --- from a
domain-specific research question to a single instantiation of a statistical
model (among many alternatives which may be subsequently explored) --- occurs.
Our depiction of hypothesis formalization aims to account for more
domain-general steps and artifacts, but we recognize that domain expertise and
norms may determine which paths and how quickly analysts move through hypothesis
formalization.

In summary, our work differs in (i) scope and (ii) method from prior work in HCI
on data analysis practices. Whereas hypothesis formalization has remained
implicit in prior descriptions of data analysis, we explicate
this specific process. While previous researchers have relied primarily on
post-analysis interviews with analysts, our lab study enables us to observe
decision making during hypothesis formalization in-situ.

\subsection{Expected Steps in Hypothesis Formalization}

Towards our first research question (\rqSteps), prior work
suggests that hypothesis formalization involves steps in three categories:
conceptual, data-based, and statistical. \textit{Conceptually,} analysts
develop conceptual hypotheses and causal models about their domain that guides
their data analysis. With respect to \textit{data}, analysts explore data and
incorporate insights from exploration, which can be top-down or bottom-up, into
their process of formalizing hypotheses. The \textit{statistical} concerns
analysts must address involve mathematical and computational concerns, such as
identifying a statistical approach (e.g., linear modeling), representing the
problem mathematically (e.g., writing out a linear model equation), and then
implementing those using software. In our work, we find
evidence to support separating statistical considerations into concerns about
mathematics, statistical specification in tools, and model implementation using
tools.

A key observation about prior work is that there is a tension between iterative and
linear workflows during hypothesis formalization. Although sensemaking processes
involve iteration, concerns about methodological soundness, as evidenced in
pre-registration efforts that require researchers to specify and follow their
steps without deviation, advocate for, or even impose, more linear processes. More
specifically, theories of sensemaking that draw on cognitive science, in
particular~\cite{russell1993cost,grolemund2014cognitive}, propose larger
iteration loops between conceptual and statistical considerations. Research
concerning statistical thinking and
practices~\cite{wild1999statisticalThinking,carver2016guidelines} appear
less committed to iteration. Empirical work in HCI on data analysis embraces
iteration during exploration and observes iteration during some phases of
confirmatory data analysis, such as statistical model choice, but not in others,
such as tool selection. 
In our work, we are sensitive to this tension and aim to provide more granular
insight into iterations and linear processes involved in hypothesis
formalization. We also anticipate that the steps identified in prior work will
recur in our content analysis and lab study, but we do not limit our
investigation to these steps. 

\section{Content Analysis of Research Papers} \label{sec:litSurvey} 

To identify what actions analysts take to formalize hypotheses (\rqSteps) and
why (\rqProcess), we examined 50 peer-reviewed publications. For breadth, we
sourced the papers from five different domains. 

\subsection{Methods}
\textbf{Dataset:} Our dataset consisted of 50 randomly sampled papers published
in 2019, ten from each of following venues: (1) the Proceedings of the National
Academy of Sciences (PNAS), (2) Nature, (3) Psychological Science (PS), (4)
Journal of Financial Economics (JFE), and the (5) ACM Conference on Human
Factors in Computing Systems (CHI). We sampled papers that used statistical analyses as either primary or secondary methodologies from these venues because
they represent a plurality of domains and Google Scholar listed them among the
top three in their respective areas in 2018.\footnote{Venues were often
clustered in the rankings without an obvious top-one, so we chose among the top
three based on ease of access to publications (e.g., open access or access
through our institution).} We analyzed published papers because researchers are
not only likely but required to report significant operationalization choices in
their publications. Finally, we focused on papers published in 2019 to capture
most recent practices.\footnote{Some papers were accepted and published before
2019, but the journals had included them in 2019 issues.}



\noindent\textbf{Analysis Approach:} When coding and analyzing the papers, we
were interested in learning about the breadth of steps involved in hypothesis
formalization rather than assessing how well papers fit a predetermined set of
steps. We also wanted to detect any co-occurrence or ordering among steps.
Based on these considerations, the first two authors iteratively developed a
codebook to code papers at the paragraph-level. The codebook contained five
broad categories: (i) research goals, (ii) data sample information, (iii)
statistical analysis, (iv) results reporting, and (v) computation. Each category
had more specific codes to capture more nuanced differences between papers. For
example, research goals could be statements or questions about something unknown,
or focused examinations of possible associations between constructs, among other
codes. This tiered coding scheme enabled us to see general content patterns
across papers and nuanced steps within papers. The first two authors reached
substantial agreement (IRR = .69 - .72) even before resolving disagreements. The
first three authors then (i) read and coded all sections of papers except the
figures, tables, and auxiliary materials;\footnote{PNAS and Nature papers
included a materials and methods section after references that were distinct
from extended tables, figures, and other auxiliary material. We coded the
materials and methods sections.} (ii) discussed and summarized the papers'
goals and main findings to ensure comprehension and identify contribution types;
and (iii) visualized and scrutinized each paper as a ``reorderable
matrix''~\cite{bertin2011graphics}. Our supplementary materials include our
codebook with definitions and examples and
summaries, citations, and matrices for each paper.

\subsection{Findings}
We coded a total of 2,989 paragraphs across 50 papers. Results were the most
commonly discussed topic. Approximately 31\% of the paragraphs (in 50 papers)
discussed interpretations of statistical results, and ~11\% (in 37 papers)
provided details about statistical results (e.g., parameter estimates).
Interpreted results often co-occurred with statistical results. ~21\% of
paragraphs (in 40 papers) described data collection design (e.g., how the
experiment was designed, how the data were collected, etc.). Specifications of
statistical models appeared in ~19\% of paragraphs (in 50 papers). ~11\% of
paragraphs (in 45 papers) discussed proxy variables, or measures to quantify
abstract constructs (e.g., music enjoyment). 

Researchers mentioned software used for statistical analysis in 3\% of
paragraphs (in 25 papers), sometimes even specifying function names and
parameters, a level of detail we did not expect to find in publications. To our
surprise, more papers mentioned software than included equations. Only fifteen
papers (JFE: 9, PS: 5, PNAS: 1) included equations in a total of 71 paragraphs.
This suggests that mathematical equations, though part of the hypothesis
formalization process, are less important to researchers than their
tool-specific implementations.

Papers published in PNAS and Nature had noticeably different structures than the
CHI, JFE, and PS papers. The PNAS and Nature papers decoupled research goals,
data sample properties, and results (in the main paper body) from data
collection design and statistical analysis (in the appended materials and
methods section). For individual studies in the CHI, JFE, and PS papers, codes
repeated in noticeably linear patterns from research goals to data collection
and sample information to proxy variables and statistical analyses to results.
We also examined papers' contribution types, identifying those that presented
empirical findings (41 papers), validated a prototype system (8 papers), or
developed a new methodology (6 papers). We include a detailed description of
paper contributions and venue differences in our supplementary material.


\theme{Researchers decompose hypotheses into sub-goals that correspond to statistical analyses.}

In approximately 70\% of papers in the corpus, we found that researchers
deconstructed their motivating research questions and overarching hypotheses into
more tightly scoped objectives or relationships of interest that map to
specific statistical analyses. For example, in~\cite{N8}, the researchers asked how theories of macroevolution
varied across groups of species. The authors divided pre-existing hypotheses
into three classes of hypotheses and assessed each class in turn. For one class
of ``geometric'' hypotheses about insect egg size, the researchers
discriminated between two opposing hypotheses by examining ``the scaling
exponent of length and width (the slope of the regression of log-transformed
length and log-transformed width).'' As this example demonstrates, hypothesis
formalization involves an \emph{iterative hypothesis refinement process at the
conceptual level}. This refinement process distills hierarchies of hypotheses
and/or a single conceptual hypothesis into sub-hypotheses and formalizes these
sub-hypotheses in statistical model implementations. Researchers also relate sub-hypotheses to one other during this process, which implies their causal models about the motivating conceptual hypothesis (and domain).


\theme{Researchers concretize hypotheses using proxies that are based on theory or available data.}
Proxy variables further refine conceptual hypotheses by identifying how observable some
concepts are, measuring the observable ones, indirectly measuring the less
observable ones, and comparing measurement choices to other possible measures or
ideal scenarios. As such, proxy variable selection is an important
transition step between conceptual and data concerns during hypothesis
formalization.

When defining proxy variables, researchers (i) used previously validated
measures when available for theoretical and methodological soundness, such as
the Barcelona Music Reward Questionnaire (BMRQ) to measure music reward
(in~\cite{PS1}), or (ii) developed new measures as a research contribution. For
example, in~\cite{CHI0}, the authors develop an EEG-based measure for
``immersiveness'' in VR they demonstrated to be superior to previous measures
that required halting immersive VR experiences to ask users about immersion.
Researchers also sometimes justified choosing proxies based on available data.
For example, in~\cite{JFE5}, the researchers wanted to develop a proxy variable
for job rank based on titles and ``financial outcomes'' (e.g., compensation,
bonuses, etc.) to see if housing bankers were promoted or demoted after the 2008 stock market
crash. However, because the financial outcomes were not public, the researchers
relied on title only to compare bankers' ranks, which was sub-optimal because
job titles differ between companies. 



Researchers consider their proxy choices as study limitations and consider
alternative proxies to ensure that their findings are robust. Validating
findings with multiple proxies suggests that hypothesis formalization can be a
\emph{recursive process}. Proxies lead to follow-up hypotheses about possible
latent measurement factors, for instance, which in turn lead to additional analyses that address
the same conceptual hypothesis. 


\theme{Data collection and sampling influence statistical analysis.}
Researchers often described their data sampling and study design as factors that
necessitated additional steps in their analysis process. In~\cite{PS0}
and~\cite{PS5}, researchers accounted for effects of task order in their study
protocol by running additional regressions or analyzing tasks separately.
Researchers also ran initial analyses to assess the effect of possibly
confounding variables in their study design, such as gender in~\cite{PS3} or
location of stimuli in~\cite{PS4}. Other times, researchers performed robustness
checks after their main analyses, such as in response to a gender imbalance
in~\cite{PS5} and possible sample selection biases due to database constraints
in~\cite{JFE1}.

Although data collection driven by statistical modeling plans was expected of
replication studies (e.g.,~\cite{PS8,PS5,PS0}) or papers that make
methodological contributions (e.g.,~\cite{JFE6, JFE7}), we found an instance
in~\cite{PS2}\textemdash neither replication nor methodological
contribution\textemdash where researchers explicitly reported selecting a
statistical model before designing their study. The researchers chose to use a
well-validated computational model, the linear ballistic accumulator (LBA), to
quantify aspects of human decision making. This model selection influenced the
way they designed their study protocol so that they could obtain a sample large
enough for accurate parameter estimation. 

Based on these observations, it seems that modeling choices more frequently
react to data collection processes and possible sample biases, following a
linear data collection-first process implied by prior work. However, there are
also instances where model implementation comes first and researchers' data
collection procedures must adhere to modeling needs, suggesting a previously
missing \emph{loop between statistical model implementation and data collection}
that is separate from any influences conceptual hypotheses have on data
collection. 

\subsection{Takeaways from the Content Analysis}
The content analysis confirmed prior findings on (i) the connection between
hypotheses and causal models, (ii) the importance of proxies to quantify
concepts, and (iii) the constraints that data collection design and logistics
place on modeling. Extending prior work, the content analysis also (i) suggested
that decomposing hypotheses into specific objectives is a mechanism by which
conceptual hypotheses relate to causal models; (ii) crystallized the hypothesis
refinement loop involving conceptual hypotheses, causal models and proxies; and (iii)
surfaced the dual-search nature of hypothesis formalization by suggesting that
model implementation may shape data collection. 

The content analysis also raised questions about how much the pressures to write
compelling scientific narratives~\cite{kerr1998harking} influence which aspects
of hypothesis formalization are described or omitted (e.g., in practice, model
implementations may constrain data collection more often than we found in our
dataset), how the steps are portrayed linearly even though the process may have been more iterative, how analysts determine
which tools to use, and how analysts without domain expertise may approach hypothesis formalization differently. These questions motivated us to design and conduct a lab
study to provide greater visibility into how analysts who are not necessarily
researchers approach the process with expectations of rigor but without pressure
of publication. 

\section{Exploratory Lab Study} \label{sec:inLabStudy}

To understand analysts' considerations (\rqProcess) while
formalizing their hypotheses (\rqSteps), as well as the role of statistical
software in this process (\rqTools), we designed and conducted a virtual lab
study.

\subsection{Methods} 
\textbf{Analysts:} We recruited 24 data analysts with experience in domains
ranging from marketing to physics to education, through Upwork (22) and by
word of mouth (2).
Twelve analysts worked as scientists, freelance data scientists,
project managers, or software engineers. Six were currently enrolled in or just
finished graduate programs that involved data analysis. Five identified as
current or recent undergraduates looking for jobs in data science. One was an
educator. Analysts self-reported having significant experience on a 10-point
scale adapted from~\cite{feigenspan2012measuring} (min=2, max=10, mean=6.4,
std=2.04) and would presumably have familiarity with hypothesis formalization.


\noindent\textbf{Protocol:} Based on our content analysis findings, we designed
and conducted a lab study with three parts. We compensated analysts \$45 for
their time. The first author conducted the study and took notes throughout.
Parts 1 and 3 were recorded and automatically transcribed using Zoom. 

\textit{Part 1: Structured Tasks.}  
Part 1 asked analysts to imagine they were leading a research team to answer
the following research question: ``What aspects of
an individual's background and demographics are associated with income after
they have graduated from high school?''\footnote{We chose the open-ended research
question about income after high school because we expected it to be widely
approachable and require no domain expertise to understand.}
We asked analysts to complete the following tasks:
\begin{itemize}
    \item \textit{Task 1: Hypothesis generation.} Imagining they had access to
    any kind of data imaginable, analysts brainstormed at least three
    hypotheses related to the research question.
    \item \textit{Task 2: Conceptual modeling.} Next, analysts saw a sample
    data schema and developed a conceptual model for one or more of their
    hypotheses. We used the term ``conceptual model'' instead of ``causal
    model'' to avoid (mis)leading analysts. We provided the following
    definition: ``A conceptual model summarizes the process by which some
    outcome occurs. A conceptual model specifies the factors you think influence
    an outcome, what factors you think do not influence an outcome, and how
    those factors might interact to give rise to the outcome.'' 
    \item \textit{Task 3: Statistical model specification.} Finally, we
    presented analysts with a sample dataset and instructed them to specify
    but not implement a statistical model to test one or more of their
    hypotheses. 
\end{itemize}

After the three tasks, we conducted a semi-structured interview with
analysts about (i) their validity concerns\footnote{If analysts were
unfamiliar with the term ``validity,'' we rephrased the questions to be about
``soundness'' or ``reliability.''} and (ii) experiences. To help us
contextualize our observations and assess the generalizability of our findings,
we asked analysts to compare the study's structure and tasks to their
day-to-day data analysis practices.

\textit{Part 2: Take-home analysis.} After the first Zoom session, analysts
implemented their analyses using the previously shown dataset, shared any
analysis artifacts (e.g., scripts, output, visualizations, etc.), and completed
a survey about their implementation experience. Prior to Part 3, the first
author reviewed all submitted materials and developed participant-specific
questions for the final interview.

\textit{Part 3: Final Interview.} The first author asked analysts to give an
overview of their analysis process and describe the hypotheses they tested, how
their analysis impacted their conceptual model and understanding, why they made
certain implementation choices, what challenges they faced (if any), and any
additional concerns about validity.

\noindent\textbf{Materials:} The data schema and dataset used in the study came from a
publicly available dataset from the Pew Research Center~\cite{pewDataset}. Each
task was presented in a separate document. All study materials are included as
supplementary material.

\noindent\textbf{Analysis:} The first author reviewed the analysts' artifacts multiple
times to analyze their content and structure;
thematically analyzed notes and transcripts from analysts' Zoom sessions;
and regularly discussed observations with the other authors throughout analysis.

\subsection{Findings and Discussion} 
Eighteen of the 24 analysts we recruited completed all three parts of the study.
The other six analysts completed only the first Zoom session. In our analysis,
we incorporate data from all analysts for as far as they completed the study. 

We found that analysts had four major steps (\rqSteps) and considerations
(\rqProcess): (i) identifying or creating proxies, (ii) fitting their present
analysis to familiar approaches, (iii) using their tools to specify models
(\rqTools), and (iv) minimizing bias by being data-based. Analysts also faced
challenges acquiring and incorporating domain and statistical knowledge
(\rqProcess).


\theme{Analysts consider proxies and data collection while articulating hypotheses.}
We encouraged analysts to not consider the feasibility of collecting data
while brainstorming hypotheses. Yet, while brainstorming hypotheses,
analysts expressed concern with how to measure constructs [A2, A5, A8,
A12, A18, A22, A24] and how to obtain data [A2, A6, A8, A9, A11, A21, A24].

For instance, A18 grappled with the idea of `privilege' and how to best quantify it:
\longquote{I'm trying to highlight the fact that those who will be privileged
before graduation...that experience will enable them to make again more money
after graduation. I won't say `privilege' because we need to quantify and
qualify for that...it's just an abstract term.} Eventually, A18 wrote two
separate hypotheses about `privilege,' operationalizing it as parental income:
(1) ``People with higher incomes pre graduating, end up having higher
differences between pre and post graduation incomes than those with lower
incomes pre graduation.'' and (2) ``People with parents with lower incomes tend
to have lower incomes pre graduation than those with parents with higher
incomes.'' 

A18 continued to deliberate `privilege' as measured by low and high income,
saying, \shortquote{...again you need to be careful with low and high because
these are just abstract terms. We need to quantify that. What does it mean to be
`low?' What does it mean to be `high?'}. Finally, A18 decided to
\shortquote{maybe use the American standards for low income and high income.}
Although an accepted ``American standard'' may not exist, A18 nevertheless
believed that cultural context was necessary to specify because it could provide
a normalizing scale to compare income during analysis, demonstrating how
analysts plan ahead for statistical modeling while brainstorming and refining
hypotheses. 

Similarly, A2 was very specific about how to measure personality:
``More extraverted individuals (extraversion measured using
the corresponding social network graph) are likely to
achieve higher yearly income later in life.'' 

In the presence of the data schema, more analysts were concerned with proxies
[A2, A5, A6, A7, A8, A9, A16, A18, A21]. Some even adapted their working
definitions to match the available data, similar to how researchers in the
content analysis determined proxies based on data. For instance, A8, who hypothesized that
``individuals interested in STEM fields tend to earn more post high school than
individuals interested in other fields,'' operationalized ``interest'' as
``Major'' --- a variable included in the data schema --- even though they had
previously brainstormed using other proxies such as club attendance in high school.


These analysts' closely related considerations of data and concept measurement
demonstrate how conceptual hypotheses and data collection may inform each other,
corroborating our findings from the content analysis.

\theme{Analysts consider implementation and tools when specifying statistical models.}
\figureLabStudyStatSpec When we asked analysts to specify their models without
considering implementation, we anticipated they would write mathematical models
that they could then implement using their tools because (a) some researchers in
the literature survey did so in their papers and (b) several analysts mentioned
having years of analysis experience. However, despite the explicit instruction
to disregard implementation, 16 analysts provided to-do lists or summaries of
steps to perform a statistical analysis as their model specifications [A1, A2,
A3, A5, A7, A8, A9, A11, A12, A14, A16, A18, A20, A21, A22, A24]. Only one
analyst, A19 (6/10 analysis experience), attempted to specify their models
mathematically but gave up because although they knew the general form of
logistic regression, they did not know how to represent the specific variables
in the model they wanted to perform. 

For example, A8 (7/10 for analysis experience), specified a list of steps that
included creating new variables that aggregated columns in the dataset, cleaning
and wrangling the data, visualizing histograms, performing chi-squared test, and
interpreting the statistical results. Notably, A8 also specified null and
alternative hypotheses, which acted as an intermediate artifact during
hypothesis formalization. Figure~\ref{figure:labStudyStatSpec} shows A8's
statistical specification.

Nine analysts went so far as to mention specific libraries, even functions, that they
would use to program their analyses [A3, A9, A12, A13, A14, A16, A19, A21,
A23]. In their reflective interviews, analysts also expressed that they often do
not specify models outside of implementing them, which A19 succinctly described:
\longquote{I don't normally write this down because all of this is in a
[software] library.} 

The implementation and software details analysts discussed and included in their
specifications suggest that analysts skip over mathematical equations and jump
to specification and implementation in their tools even though some papers
included equations as an intermediate step in our content analysis.
Additionally, analysts' statistical knowledge is situated in the programs they
write, and their knowledge of and familiarity with tools constrains the
statistical methods they explore and consider. As such, tools may be a key point
of intervention for guiding analysts toward statistical methods that may be
unfamiliar but are best suited for their conceptual hypotheses.

\theme{Analysts try to fit analyses to previous projects and familiar approaches.}
Analysts spent significant thought and time categorizing their analyses as
``prediction'', ``classification'' or ``correlation'' problems [A2, A3, A7,
A10, A11, A18, A19, A21, A22]. To categorize, analysts relied on their
previous projects. While reflecting on their typical analysis process, A21 said,
\longquote{I usually tend to jump...to look at data and \textbf{match [the
analysis problem] with similar patterns} I have seen in the past and start
implementing that or do some rough diagrams [for thinking about parameters, data
type, and implementation] on paper...and start implementing it.} 

Analysts also looked at variable data types (i.e., categorical or continuous) to
categorize. For example, A3 pivoted from thinking about \textbf{predicting}
income to \textbf{classifying} income groups (emphasis added) based on data type
information: \longquote{The income, the column, the target value here, is
categorical. I think maybe it wouldn't be a bad idea to see what
\textbf{classification} tasks, what we could do. So instead of trying to
\textbf{predict} because we're not trying to \textbf{predict an exact number},
it seems...like more of a \textbf{classification} problem...}

A provocative case of adhering to prior experiences was A6. Although several
analysts were surprised and frustrated that income was ordinal in the dataset
with categories such as``Under \$10K,'' ``\$10K to \$20K,'' ``\$20K to \$30K,''
up to ''150K+'', none went so far as A6 to synthetically generate normally
distributed income data so that they could implement the linear regression
models they had specified despite saying they knew that income was not normally
distributed. 

When asked further about the importance of normal data, A6, a research
scientist, described how they plan analyses based on having normal data, strive
to collect normally distributed, and rely on domain knowledge to transform the
data to be normal when it may not be after collection: \longquote{...I feel like
having non normal data is something that's like hard for us to deal with. Like
it just kind of messes everything up like. And I know, I know it's not always
assumption of all the tasks, but just that we tend to try really hard to get our
variables to be normally distributed. So, you know, we might like transform it
or, you know, kind of clean it like clean outliers, maybe transform if
needed...I mean, it makes sense because like a lot of measures we do use are
like depressive symptoms or anxiety symptoms and kind of they're naturally
normally distributed...I can probably count on my hand the number of non
parametric tests I've like included in manuscripts.} A6's description of their
day-to-day analyses exemplifies the dual-search nature of hypothesis
formalization: Analysts (i) jump from hypothesis refinement to model
specification or implementation with specific proxies in mind and then (ii)
collect and manipulate their data to fit their model choices. 

We recognize that analysts may have taken shortcuts for the study they would
not typically make in real life. Nevertheless, the constraints we imposed by
using a real-world dataset are to be expected in real-world analyses. Therefore,
our observations still suggest that rather than consider the nature and
structure of their hypotheses and data to inform using new statistical
approaches, which statistical pedagogy and theory may suggest, analysts may
choose familiar statistical approaches and mold their new analyses after
previous ones. 


\theme{Analysts try to minimize their biases by focusing on data.}

Throughout the study, analysts expressed concern that they were biasing the
analysis process. Analysts drew upon their personal experiences to develop
hypotheses [A5, A10, A13, A15, A16, A20, A21, A24] and conceptual models [A8,
A12, A20, A24]. A12 described how their personal experiences may subconsciously
bias their investigation by comparing a hypothetical physicist and social worker
answering the same research question: \longquote{Whereas a social worker by
design...they're meant to look at the humanity behind the numbers [unlike a
physicist]. So like, they may actually end up with different results...actually
sitting in front of this data, trying to model it.}

A few analysts even refused to specify conceptual models for fear of biasing the
statistical analyses [A10, A11, A19]. On the surface, analysts resisted
because they believed that some relationships, such as the effect of age on
income, were too ``obvious'' and did not warrant documentation [A10, A11].
However, relationships between variables that were ``obvious'' to some
analysts were not to others. For instance, A10 described how income would
plateau with age, but other analysts, such as A18 while implementing their
analyses, assumed income would monotonically increase with age. 

When we probed further into why A10, A11, and A19 rejected a priori conceptual
models, they echoed A10's belief that conceptual models ``put blinders on you.''
Even the analysts who created conceptual models echoed similar concerns of
wanting to ``[l]et the model do the talking'' in their implementations [A3, A15,
A18, A19]. Instead of conceptual modeling, A10 chose to look at all n-ary
relationships in the dataset to determine which variables to keep in a final
statistical model, saying, \longquote{It's so easy to run individual tests...You
can run hypothesis tests faster than you can actually think of what the
hypothesis might be so there's no need to really presuppose what relationships
might exist [in a conceptual model].} Of course, one could start from the
same premise that statistical tests are so easy to execute and conclude that
conceptual modeling is all the more important to prioritize analyses
and prevent false discoveries. 

Similarly, analysts were split on whether they focused their implementation exclusively on their
hypotheses or examined other relationships
in the dataset opportunistically. Nine analysts stuck strictly to testing their hypotheses [A1,
A4, A5, A6, A7, A11, A13, A20, A24]. However, five analysts were more focused on
exploring relationships in the dataset and pushed their hypotheses aside [A2,
A3, A10, A16, A18], and an additional four analysts explored relationships among
variables not previously specified in their hypotheses in addition to their
hypotheses [A14, A15, A17, A21]. A18 justified their choice to ignore their
hypotheses and focus on emergent relationships in the data by saying that they
wanted to be \shortquote{open minded based on the data...open to possibilities.}

Analysts' concerns about bias and choice of which relationships to analyze
(hypothesis only vs. opportunistic) highlight the tension between the two
searches involved in hypothesis formalization: concept-first model
implementations and implementation-first conceptual understanding. Conceptual
models are intermediate artifacts that could reconcile the two search processes
and challenge analysts' ideas of what ``data-driven'' means. However, given some
analysts' resistance to prior conceptual modeling, workflows that help
analysts conceptually model as a way to reflect on their model implementations
and personal biases may be more promising than ones that require them before
implementation.

\theme{Analysts face challenges obtaining and integrating conceptual and statistical information.}
Based on analysts' information search behaviors and self-reports, we found that
analysts faced challenges obtaining and integrating both domain and statistical
knowledge.

Analysts consulted outside resources such as API documentation, Wikipedia, and
the \textit{Towards Data Science} blog throughout the study: one while
brainstorming hypotheses [A13]; three while conceptual modeling [A12, A13, A22];
six while specifying statistical models [A3, A6, A12, A13]. Six analysts
also mentioned consulting outside resources while implementing their analyses
[A1, A3, A11, A14, A15, A21]. By far, statistical help was the most common. 

Furthermore, when analysts reflected on their prior data analysis experiences,
they detailed how collaborators provided domain and statistical expertise that
are instrumental in formalizing hypotheses. Collaborators share data that help
domain experts generate hypotheses [A9], critique and revise conceptual models
and proxies [A4, A8], answer critical data quality questions [A10],
and ensure statistical methods are appropriate [A5, A6, A22].

In the survey participants completed after implementing their analyses, the three most
commonly reported challenges were (i) \textbf{formatting} the data [A1, A4,
A5, A6, A13, A16, A18, A20, A21, A24], (ii) \textbf{identifying} which
statistical analyses to perform with the data to test their hypotheses [A1,
A11, A14, A18, A20, A21], and (iii) \textbf{implementing and executing} analyses
using their tools [A1, A6, A7, A13, A20, A21]. Although we expected analysts
would have difficulty wrangling their data based on prior
work~\cite{kandel2012enterprise}, we were surprised that identifying and
executing statistical tests were also prevalent problems given that (a) analysts
were relatively experienced and (b) could choose their tools. These results, together with 
our observations that analysts rely on their prior experiences and tools, suggest
that analysts have difficulty adapting to new scenarios where new tools and
statistical approaches may be necessary. 

\subsection{Takeaways from the Lab Study}
After the first session, 13 out of the 24 analysts described all the tasks as
familiar, and 10 described most of the tasks and process as familiar. Analysts
commonly remarked that although the process was familiar, the order of the tasks
was ``opposite'' of their usual workflows. In practice, analysts may start with
model implementation before articulating conceptual hypotheses, which opposes
the direction of data analysis that the ASA
recommends~\cite{carver2016guidelines}. Nevertheless, our observations reinforce
the dual-search, non-linear nature of hypothesis formalization.

Moreover, one analyst, A24, a physics researcher who primarily conducted
simulation-based studies expressed that the study and its structure felt
foreign, especially because they had no control over data collection. Other
analysts in the study also described the importance of designing and conducting
data collection as part of their hypothesis formalization process [A4, A6, A9].
Designing data collection methods informs the statistical models analysts plan
to use and helps to refine their conceptual hypotheses by requiring analysts to
identify proxies and the feasibility of collecting the proxy measures,
reinforcing what we saw in the content analysis. The remarks also suggest that
disciplines practice variations of the hypothesis formalization process we
identify based on discipline-specific data collection norms and constraints. For
example, simulating data may sometimes take less time than collecting human
subjects data, so analysts working with simulations may dive into modeling and
data whereas others may need to plan experiments for a longer period of time. 

Finally, we found that analysts relied on prior experiences and tools to specify
and formalize their hypotheses. Tools that scaffold the hypothesis formalization
process by suggesting statistical models that operationalize the conceptual
hypotheses, conceptual models, or partial specifications analysts create along
the way may (i) nudge analysts towards more
robust analyses that test their hypotheses, (ii) overcome limitations of analysts'
prior experiences, and (iii) even expand analysts' statistical knowledge. Thus, we
investigated how current tool designs serve (or under-serve) hypothesis
formalization.

\section{Analysis of Software Tools} \label{sec:toolsAnalysis}

To understand how the design of statistical computing tools may support or
hinder hypothesis formalization (\rqTools), we analyzed widely
used software packages and suites. Our observations in the lab study motivated
and focused our analysis. 
Throughout, we use the term ``package`` to refer to a set of programs that must be invoked
through code, such as \texttt{lme4}, \texttt{scipy}, and \texttt{statsmodels}. We use the term ``suite``
to refer to a collection of packages that end-users can access either through
code or graphical user interfaces (GUIs), such as SPSS, SAS, and JMP. We use
the term ``tool'' to refer to both.

\subsection{Method}

\textbf{Sample:} We consulted online data analysis
fora~\cite{grolemund2019:recommendedR, bobriakov2017:top15Python,
bobriakov2018:top20Python, prabhu2019:topPython} to identify and include widely
used tools. We also included tools that appeared more than once across the
content analysis and lab study. The final sample included 20 statistical tools:
14 packages (R: 10, Python: 4); three suites that support in-tool programming;
and three suites that do not support programming.
Table~\ref{tableAnalysisOfTools} contains an overview of our sample and results. 

\noindent\textbf{Analysis:} Four specific questions guided our analysis:
\begin{itemize}
    \item \textbf{Specialization:} Analysts in the lab study eagerly named
    specific statistical tools they would use and looked up tool documentation
    during the tasks. This prompted us to ask \textit{How specialized are the
    tools, and how might specialization (or lack thereof) affect how analysts
    discover and use them to formalize hypotheses?}
    \item \textbf{Statistical Taxonomies:} Analysts in the lab study tried to
    mold their analysis to prior experiences and their taxonomies of statistical
    methods. We wondered what role tools play in this: \textit{How do tools
    organize and group statistical models? How might tool organization and
    analysts' taxonomies interplay during hypothesis formalization?}
    \item \textbf{Model Expression:} Analysts in the lab study did not specify
    their statistical models mathematically. Instead, they jumped to model
    implementation. We wondered if this was due to how tools enable analysts to
    express their models: \textit{What notation must analysts use to express
    models in the tools?}
    \item \textbf{Computational Issues:} It was uncommon for researchers in the
    content analysis to specify the software tools they used. However, when
    researchers did mention software, they also specified the functions,
    parameters, and settings used, prompting us to wonder about the importance
    of computational settings: \textit{What specific kinds of computational
    control do tools provide end-users and how might that impact hypothesis
    formalization?}
\end{itemize}

To answer the four questions for each statistical tool, the first author read
and took notes on published articles about tools' designs and implementations,
API documentation and reference manuals, and available source code; followed
online tutorials; consulted question-and-answer sites (e.g., StackExchange) when
necessary; and analyzed sample data with the tools. The first author paid
particular attention to tool organization, programming idioms, functions and
their parameters, and tool failure cases. Table~\ref{tableAnalysisOfTools}
contains citations for resources consulted in the analysis. 

The analysis process was iterative and involved discussions among the co-authors
about how to evaluate the properties of tools from our perspectives as both tool
designers/maintainers and end-users. Here, we focus on end-user (hereafter
referred to as analyst) perspectives informed by our lab study and make callouts
to details relevant for tool designers.

\subsection{Findings and Discussion}
\tableSoftwareAnalysis
We discuss our findings in light of our characterization of hypothesis
formalization in Figure~\ref{figure:overview}. We refer to specific steps and
transitions in Figure~\ref{figure:overview} in \textbf{boldface}.


\theme{Tool specialization pushes computational concerns higher up the hypothesis formalization process.}
Half the tools [T2, T3, T4, T5, T6, T7, T8, T9, T11, T12] in our sample are
specialized in the scope of statistical analysis methods they support (e.g.,
\texttt{brms} supports Bayesian generalized linear multilevel modeling).
\texttt{edgeR} [T3] provides multiple modeling methods but is specialized to the
context of biological count data. Such specialized tools are vital to creating a
widely adopted statistical computing ecosystem, such as R. However, highly
specialized packages also come at a cost. Specialized tools require analysts to
consider computational settings while picking a statistical tool and, possibly,
even while mathematically relating their variables. Thus, specialized tools fuse the last two steps of hypothesis formalization
(\textbf{Statistical Specification} and \textbf{Model Implementation}).
Ultimately, specialization requires analysts to have more (i) computational
knowledge and (ii) foresight about their model implementations at the cost of
focusing on conceptual or data-related concerns early in hypothesis
formalization. 



One way tool designers minimize the requisite computational knowledge and
foresight while providing the benefits of specialized packages --- which may be
optimal for specific statistical models or data analysis tasks --- is to provide
micro-ecosystems of packages. For example, R's
\texttt{tidymodels}~\cite{tidymodels} and \texttt{tidyverse}~\cite{tidyverse}
create micro-ecosystems that use consistent API syntax and semantics across
interoperable packages. They also push analysts towards what the tool designers
believe to be best practices, such as the use of the tidy data format. Tools
that aim to support hypothesis formalization may consider fitting into or
creating micro-ecosystems that provide tool support all along the process, focusing analysts on concepts, data, or model
implementation at various points. 

\theme{Tool taxonomies introduce challenges that detract from hypothesis
formalization.} 

A consequence of tool specialization is the fragmented view of statistical
approaches. For example, we observed analysts in the lab study who viewed the
analysis as a classification task gravitate towards machine learning-focused
libraries, such as \texttt{RandomForest} [T9], \texttt{Keras} [T11], and
\texttt{scikit-learn} [T12]. Because classification can be implemented as
logistic regression, any tool that supports logistic regression, such as the
core \texttt{stats} library in R [T10], provides equally valid, alternative
perspectives on the same analysis and hypothesis. However, tools obfuscate these
connections and do not aid analysts in considering reasonable statistical models
that may be unfamiliar or outside their personal taxonomy. This may explain why
analysts adhered to their personal taxonomies during the lab study.

This problem carries over to tools that support numerous statistical methods.
Ten tools in our sample intend to provide more comprehensive statistical
support [T1, T10, T13, T14, T15, T16, T17, T18, T19, T20]. These tools group
statistical approaches using brittle and inconsistent taxonomies based on data
types [T17]; tool-defined analysis classes that are both highly specific
(e.g., ``Item Response Theory'') and vague (e.g., ``Multivariate analyses'')
[T15, T16, T17, T18, T19, T20]; and disciplines or applications (e.g.,
``Epidemiology and related,'' ``Direct Marketing'') [T16, T17, T20]. Although
well-intended to simplify statistical method selection, tools' taxonomies are at
times misleading. For instance, JMP combines various linear models into a ``Fit
Model'' option that is separate from ``Predictive Modeling'' and ``Specialized
Modeling,'' which are also distinct from the more general ``Multivariate
Methods.'' Once analysts select the ``Fit Model'' option, they can specify the
``Personality'' of their model as ``Generalized Regression,'' ``Generalized
Linear Model,'' or ``Partial Least Squares,'' among many others. This JMP menu structure may lead an analyst to conclude that (i) a Partial Least Squares model is
distinct from a regression model when it is in fact a type of regression model
and (ii) regression is not useful for prediction. 

In these ways, tools add a ``Navigate taxonomies'' step before the
\textbf{Statistical Specification} step, requiring analysts to match their
conceptual hypotheses with the tools' taxonomies, which may misalign with their personal taxonomies. One reason for this issue may be that tools do
not leverage analysts' intermediate artifacts or understanding during hypothesis
formalization. By the time analysts transition to \textbf{Statistical
Specification}, they have refined their conceptual hypotheses, developed causal
models, and made observations about data. However, tools' taxonomies require
analysts to set these aside and consider another set of decisions imposed by
tool-specific groupings of statistical methods.

\theme{Syntactic and semantic mismatches create a rift between model implementations and conceptual hypotheses.}

Fifteen tools in our sample provide analysts with interfaces that use
mathematical notation to express statistical models [T1, T2, T3, T4, T6, T7, T8,
T9, T10, T14, T16, T17, T18, T19, T20]. R and Python packages use symbolic
mathematical syntax, and SPSS and Stata use natural language-like syntax.
Expressing a linear model with Sex, Race, and their interaction as predictors of
Annual Income involves the formula \texttt{AnnualIncome $\sim$ Sex + Race +
Sex*Race} in \texttt{lme4} and \texttt{AnnualIncome BY Sex Race Sex*Race} in
SPSS.  In a linear execution of steps involved in hypothesis formalization where
analysts relate variables mathematically (\textbf{Mathematical Equation}) before
specifying and implementing models using tools (\textbf{Statistical
Specification}, \textbf{Model Implementation}), the mathematical interfaces
match analysts' progression. However, in the lab study analysts did not specify
their models mathematically even when given the opportunity, suggesting that
mathematical syntax may not adequately capture analysts' conceptual or
statistical considerations. 


Syntactic similarity between packages may lower the barrier to trying and
adopting new statistical approaches but may introduce unmet expectations of
semantic similarity. For example, \texttt{brms} [T2] uses the same formula
syntax as \texttt{lme4} [T6], smoothing the transition between linear modeling
and Bayesian linear modeling for analysts. However, based on syntactic
similarity, analysts may incorrectly assume statistical equivalence in computed
model values. For example, in \texttt{brms}, the model intercept is the mean of
the posterior when all the independent variables are at \textit{their means},
but in \texttt{lme4}, the intercept is the mean of the model when all the
independent variables are \textit{at zero}. 


Furthermore, tools introduce syntactic differences between statistical
approaches that are for the most part semantically equivalent. For instance,
an ANOVA with repeated measures and a linear mixed effects model are similar in
intent but require two different function calls, one without a formula (e.g.,
\texttt{AnovaRM} in \texttt{statsmodels} [T14]) and another with (e.g.,
\texttt{mixedlm} in \texttt{statsmodels} [T14]). Even when considering only ANOVA, tools may provide similar syntax but implement different sums of squares procedures for partitioning variance (i.e., Type I, Type II, or
Type III).\footnote{Type I is (a) sensitive to the order in which independent
variables are specified because it assigns variance sequentially and (b) allows
interaction terms. Type II (a) does not assign variance sequentially and (b)
does not allow interaction terms. Type III (a) does not assign variance
sequentially and (b) allows interaction terms. For an easy-to-understand blog
post, see~\cite{sumsofsquaresBlog}.} By default, R's \texttt{stats} [T10] core
package [T10] uses Type I, \texttt{statsmodels} [T14] uses Type II, and
\texttt{SPSS} [T16] uses Type III. The three different sum of squares
procedures lead to different F-statistics and p-values, which may lead
analysts to different conclusions. More importantly, the procedures encode
different conceptual hypotheses. If analysts have theoretical knowledge or
conceptual hypotheses about the order of independent variables, tools defaulting
to Type I (e.g., R's \texttt{stats} core library) align the model
implementation with the conceptual hypotheses. However, if analysts do not have
such conceptual hypotheses, tools' default behavior would execute (without error)
and silently respond to a conceptual hypothesis different from the one the analyst
seeks to test. 

The impact of tools' ``invisible'' model implementation choices
reinforces the interplay between conceptual and model implementation concerns
during hypothesis formalization. Tools should highlight the conceptual
assumptions and consequences of modeling choices beyond listing ways to change
defaults in their documentation manuals or Q\&A sites. Doing this would give
analysts more control over and insight into their analysis. Analysts could
revise and refine their hypotheses in response to statistical modeling
constraints or change the statistical models and tools they use to address their
hypotheses. 

\theme{Fine-grained computational control may require conceptual hypothesis revisions.}
Tools provide end-users with options for optimizers and solvers used to fit
statistical models [T1, T2, T4, T6, T7, T8, T10, T11, T13, T16, T18],
convergence criteria used for fitting
models [T3, T6, T16, T18], and memory and CPU allocation [T2, T5, T12, T15], among more specific
customizations. 
For instance, \texttt{lme4} [T6] allows analysts to specify the nonlinear
optimizer and its settings (e.g., the number of iterations, convergence
criteria, etc.) used to fit models. In \texttt{brms} [T2], analysts can also
specify the number of CPUs to dedicate to fitting their models. Some
computational settings are akin to performance optimizations, affecting computer
utilization but not the results. However, not all computational changes are so
well-isolated.

For example, the lack of model convergence (in \textbf{Model Implementation})
may prompt mathematical re-formulation (\textbf{Mathematical Equation}), which
may cast \textbf{Observations about Data} in a new light, prompting
\textbf{Causal Model} and \textbf{Conceptual Hypothesis} revision. In other
words, computational failures and decisions may bubble up to conceptual hypothesis
revision and refinement, which may then trickle back down to model
implementation iteration, and so on. In this way, computational
control can be another entry into the dual-search process of hypothesis formalization. 

Although in theory this low-level control could help analysts formalize
nuanced conceptual hypotheses in diverse computational environments, we found
that tools do not currently provide feedback on the ramifications of these
computational changes, introducing a gulf of evaluation~\cite{norman1986cognitive}. Analysts can
easily change parameters to fine-tune their computational settings, but how they
should interpret their model implementations and revisions conceptually is
unaddressed, suggesting opportunities for future tools to bridge the conceptual
and model implementation gap.

\subsection{Takeaways from the Analysis of Tools}


Taken together, our analysis shows that tools can support a wide range of
statistical models but expect analysts to have more statistical expertise than
may be realistic. They provide limited guidance for analysts to identify
reasonable models and little-to-no feedback on the conceptual ramifications of
model implementation iterations. This reveals a misalignment between analysts'
hypothesis formalization processes and tools' expectations and design. Possible
reasons for this mismatch may be that tools do not scaffold or embody the
dual-search nature of hypothesis formalization or leverage all the intermediate
artifacts analysts may create (e.g., refined conceptual hypotheses, causal
models, data observations, partial specifications, etc.) throughout the process.

\section{Implications} \label{sec:implications}
Our findings and characterization of hypothesis formalization suggest three
opportunities for tools to facilitate the dual-search process and align
conceptual hypotheses with statistical model implementations at various stages
of hypothesis formalization. 


\subsection{Meta-libraries: Connecting Model Implementations with Mathematical Equations}

Specialized tools, although necessary for statistical computation, require a
steep learning curve. Additionally, how tool interfaces, model
implementation, and computed outputs are connected is
ambiguous, even if analysts use symbolic formulae.
\textit{Meta-libraries} could allow analysts to specify their models in high-level code; find the
appropriate library or libraries to execute such a model in its knowledge base;
execute the model using the appropriate libraries; and then output library
information, functions invoked, any computational settings used, the 
mathematical model that is approximated, and the model results.
For example, libraries such as Parsnip~\cite{parsnip} have begun to provide a unified
higher-level interface that allows analysts to specify a statistical model using
more ``generically'' named functions, parameter names, and symbolic formulae
(when necessary). Parsnip then compiles and invokes various library-specific
functions for the same statistical model.

Meta-libraries
designed as such could bring three benefits. First, they would provide simpler,
less fragmented interfaces to analysts while continuing to take advantage of
tool specialization. Second, meta-libraries that output complete mathematical
representations would more tightly couple mathematical representations with
implementations, providing an on-ramp for analysts to expand their statistical
knowledge. Third, meta-libraries that showed the mathematical representations
alongside underlying libraries' function calls could show syntactical variation
in underlying libraries, indirectly teaching analysts how they might express
their statistical models in other tools and familiarizing analysts with new
tools and models. 

Future meta-libraries could consider providing a higher-level, declarative
interface that does not require analysts to write symbolic formulae. Designing
such declarative meta-libraries would require formative elicitation studies
(similar to natural programming studies such as~\cite{verou2018extending}) on
declarative primitives that are memorable, distinguishable, and reliably
understood. An additional challenge would lie in maintaining support for various
libraries executed under the hood, especially as libraries change their APIs,
which would strengthen the case for meta-libraries. Although meta-libraries
would not solve the problems involved in understanding how computational
settings affect model execution or conceptual hypotheses, they could
nevertheless provide scaffolding for analysts to more closely examine specific
libraries, especially if multiple libraries execute the same model but do not
all encounter the same computational bottlenecks. 

\subsection{High-level Libraries: Expressing Conceptual Hypotheses to Bootstrap Model Implementations}

A possible explanation for why analysts in the lab study started with model
implementation is the absence of tools for directly expressing conceptual
hypotheses. High-level libraries could allow analysts to specify data collection
design (e.g., independent variables, dependent variables, controlled effects,
possible random effects); variable data types; expected or known covariance
relationships based on domain expertise; and hypothesized findings in a
library-specific grammar. High-level libraries could compile these conceptual
and data declarations into weighted constraints that represent the applicability
of various statistical approaches, in a fashion similar to
Tea~\cite{jun2019tea}, a domain-specific language for automatically selecting
appropriate statistical analyses for common hypothesis tests. Libraries could
then execute the appropriate statistical approaches, possibly by using
a meta-library as described above. 

In addition to questions of how to represent a robust taxonomy of statistical
approaches computationally, another key challenge for developing high-level
libraries is identifying a set of minimal yet complete primitives that are
useful and usable for analysts to express information that is usually expressed
at different levels of abstraction: conceptual hypotheses, study designs, and
possibly even partial statistical model specifications. For instance, even if a
conceptual hypothesis is expressible in a library, it may be impossible to
answer with a study design or partial statistical model that is expressed in the
same program. An approach may be to draw upon and integrate aspects from
existing high-level libraries and systems that aim to address separate steps of
the hypothesis formalization process, such as
Touchstone2~\cite{eiselmayer2019touchstone2} for study design and Tea and
Statsplorer~\cite{wacharamanotham2015statsplorer} for statistical
analysis. 



\subsection{Bidirectional Conceptual Modeling: Co-authoring Conceptual Models and Model Implementations}
Conceptual, or causal, modeling was difficult for the analysts in the lab study.
Some even resisted conceptual modeling for fear of biasing their analyses. Yet,
implicit conceptual models were evident in the hypotheses analysts chose to
implement and the sub-hypotheses researchers articulated in the content
analysis. 


Mixed-initiative systems that make explicit the connection between conceptual
models and statistical model implementations could facilitate hypothesis
formalization from either search process and allow analysts to reflect on their
analyses without fear of bias. For example, a mixed-initiative programming
environment could allow analysts to write an analysis script, detect data
variables in the analysis scripts, identify how groups of variables co-occur in
statistical models, and then visualize conceptual models as graphs where the
nodes represent variables and the edges represent relationships. The
automatically generated conceptual models would serve as templates that analysts
could then manipulate and update to better reflect their internal conceptual
models by specifying the kind of relationship between variables (e.g.,
correlation, linear model, etc.) and assigning any statistical model roles
(e.g., independent variable, dependent variable). As analysts update the visual
conceptual models, they could evaluate script changes the system proposes.
In this way, analysts could externally represent their causal models while
authoring analysis scripts and vice versa. 

Although bidirectional programming environments already exist for vector graphics
creation~\cite{hempel2019sketch}, they have yet to be realized in mainstream data analysis
tools. To realize bidirectional, automatic conceptual modeling, researchers would
need to address important questions about (i) the visual grammar, which would
likely borrow heavily from the causal modeling literature; (ii) program analysis
techniques for identifying variables and defining co-occurrences (e.g., line-based
vs. function-based) in a way that generalizes to multiple statistical libraries;
and (iii) adoption, as analysts who may benefit most from such tools (likely
domain non-experts) may be the most resistant to tools that limit the number of
``insights'' they take away from an analysis. 

\section{Discussion and Limitations}
Hypothesis formalization is a dual-search process of translating
conceptual hypotheses into statistical model implementations. Due to constraints imposed by domain expertise, data,
and tool familiarity, the same conceptual hypothesis may be formalized into
different model implementations. The same model implementation may also
formalize two possibly opposing hypotheses. To navigate these constraints,
analysts use problem-solving strategies characteristic of the larger scientific
discovery process~\cite{klahr1988dual,schunn1995FourSpace}. 
 
At a conceptual level, hypothesis formalization involves \textit{hypothesis refinement}, which, to use
Schunn and Klahr's language~\cite{schunn1995FourSpace}, is a \textit{scoping}
process. In the content analysis, we found that researchers \textit{decomposed}
their research goals and conceptual hypotheses into specific, testable
sub-hypotheses and \textit{concretized} constructs using proxies, born of theory
or available data. Corroborating these findings, we found that analysts in the
lab study also quickly converged on the need to specify established proxies or
develop them based on the data schema presented. In hypothesis formalization,
scoping incorporates domain- and data-specific observations to qualify the
conceptual scope of researchers' hypotheses. In other words, hypothesis
refinement is an instance of \textit{means-end
analysis}~\cite{newell1972humanProblemSolving}, a problem-solving strategy that
aims to recursively change the current state of a problem into sub-goals (i.e.,
increasingly specific objectives) in order to apply a technique (i.e., a
particular statistical model) to solve the problem (i.e., test a hypothesis). 

At the other computational endpoint of hypothesis formalization, \textit{model implementation}
also involves iteration. Through our analysis of software tools, we
found that analysts must not only select tools among an array of specialized and
general choices but also navigate tool-specific taxonomies of statistical
approaches. These tool taxonomies may both differ from and inform analysts'
personal categorizations, potentially explaining why analysts in our lab study
relied on their personal taxonomies and tools. Based on their prior experience, analysts engage in
\textit{analogical} reasoning~\cite{holland1989induction}, finding parallels between the present analysis
problem's structure and previously encountered ones or ones that fit a tool's
design easily.

Upon selecting a statistical function, analysts may tune computational settings,
choose different statistical functions or approaches, which they may tune, and
so on. In this way, the model implementation loop in hypothesis
formalization captures the ``debugging cycles'' analysts encounter, such as the census
researcher in the introduction. The tool ecosystem as a
whole supports diverse model implementations, even for the same
mathematical equation. However, the tool interfaces provide low-level abstractions, such as
interfaces using mathematical formulae that, based on our observations in the
lab study, do not support the kind of higher-level conceptual reasoning required
of hypothesis formalization.


The steps, considerations, and strategies we have identified are domain-general.
Domain-specific expertise likely influences how quickly analysts switch between
steps and strategies during the dual-search process. Domain experts, including
researchers in our content analysis, may know which statistical model
implementations and computational settings to use a priori and design their
studies or specify their conceptual hypotheses in light of these expectations
--- incorporating means-end analysis and analogical reasoning strategies ---
more quickly. It may be these insights that analysts in our lab study sought
when they looked online for conceptual and statistical help. Future work could
observe how domain experts perform hypothesis formalization and characterize
when and how analysts draw upon their own or collaborators' expertise to circumvent iterations or justify
early scoping decisions. These insights may also shed light on how
pre-registration expectations and practices could be made more effective. Given
the level of detail required of some pre-registration policies, researchers
likely engage in a version of the hypothesis formalization process we have
identified prior to registering their studies. Knowing how pre-registration fits
into the hypotheis formalization process could improve the design and adoption
of pre-registration practices.


Finally, our findings suggest opportunities for future tools to bridge steps
involved in hypothesis formalization and guide analysts towards reasonable model
implementations. Possibilities include connecting model implementations to their
mathematical representations through meta-libraries, providing higher-level
abstractions for more directly expressing conceptual hypotheses, and supporting
automated conceptual modeling. Follow-up user studies in the lab and
observational settings as well as system development and deployment are necessary to test these
ideas, draw stronger conclusions about how tool design affects hypothesis
formalization, and more readily support analysts translate their conceptual
hypotheses into statistical model implementations.




\balance{}

\bibliographystyle{SIGCHI-Reference-Format}
\bibliography{lm_hypothesis_paper.bib}

\end{document}


\maketitle

\section{Supplementary Information about Content Analysis}

We scraped the five venues' proceedings using Helena~\cite{chasins2018rousillon}
and wrote additional scripts to randomly sample from each venue. The first
author read papers and included them in the sample if they used statistical
analyses. We did not discriminate between papers that used statistical analyses
as a primary or secondary methodology. We anticipated that authors would
describe hypothesis formalization in both cases.

\subsection{Coding Procedure}
Table~\ref{table:litSurveyCodeBook} contains our codebook.

Based on exploratory rounds of open coding an on independent sample and
noticeable differences in writing structure and style across the venues, we read
all sections of papers unlike\cite{mcdonald2019reliability} who performed a
similar content analysis of qualitative analysis methodologies and focused on
methods sections. We read but excluded any figures, tables, and auxiliary
materials from our analysis for all venues except PNAS and Nature. Papers in
these venues included a materials and methods section after references that were
distinct from extended tables, figures, and other auxiliary material. 

We coded the papers at the paragraph-level. We initially started by coding at
the sentence-level but found that paragraphs provided necessary context for
accurately interpreting and coding sentences, showed co-occurrence patterns, and
were more expedient and anecdotally more reliable to code. Nonetheless,
throughout the coding process, we deliberated and discussed key sentences in
paragraphs that shaped the paper's argumentation structure. The codebook
contains such key sentences. 

After the first three authors coded, reviewed (for coding consistency), and
discussed each paper, we created ``reorderable
matrices''~\cite{bertin2011graphics} for each paper. Scrutinizing the matrices
and referring to the papers back and forth, a set of visual patterns emerged.
The visual patterns brought into focus paper content and structure that guided
our follow-up analyses and discussions about what steps (\rqSteps) and
considerations (\rqProcess) researchers reported having during hypothesis
formalization. The definitions and notes on the patterns we used are included as
supplementary material. Please refer to the README for file names and descriptions. 


\subsection{Contribution types}
After reading and coding the papers, we re-read assigned each paper at least one
of the following contribution types: Methodology, System or Technique, and
Empirical Findings, and Other. Methodological contributions introduce a new way
of measuring a concept and may be in the form of novel experimental designs,
procedures, proxies, or other measures. System or technique contributions
develop a prototype tool, which may be physical, biological, or chemical in
nature. Empirical findings contributions primarily show or explain a new
phenomenon, which may involve developing new causal models of a domain. Other
contributions included replication studies and other results that were unique to
one or two papers in our sample, such as finding a new species in~\cite{N1}. We
identified these four contribution types through discussions and open coding. 

We found that 41 papers that made empirical contributions describing or
explaining a phenomenon; eight papers that developed and evaluated physical or
biological prototype tools; and six papers that presented novel methodologies
such as experimental protocols or measures. Ten papers made various other
contributions (e.g., replicating a previous study, finding a new species,
developing a design space, etc.). Tables~\ref{table:CHIContribs}
through~\ref{table:PSContribs} give an overview of contribution types in each
venue. We separated the tables by venue due to spacing constraints. 

Papers contributing empirical findings consisted of ten papers from PNAS, ten from
PS, eight from JFE, eight from Nature, and five from CHI. Six of the eight
system/technique contributions came from CHI papers, with one each from Nature and PNAS.
Out of the six methodology contributions, three came from JFE papers, two from
CHI, and one from PNAS. Thirteen papers fell under multiple contribution types.
Co-occurrences of two out of the three contribution types were seen in a few of
the CHI and PNAS papers, with system/technique contributions co-occurring
with either methodology or empirical findings. Co-occurrences in the PS and PNAS
papers involved an "Other" contribution type occurring most often with empirical
findings. We identified only one JFE paper with multiple contributions; in this
case, methodology and empirical findings co-occurred. We did not notice any
obvious differences in paper content or structure due to research contribution
types, either within or across venues.

\clearpage
\tableLitSurveyCodes
\tableContributions

\clearpage
\bibliographystyle{../SIGCHI-Reference-Format}
\bibliography{../tea-paper,../lit_survey_papers}